\documentstyle[preprint,aps]{revtex}

\def\eps{\epsilon}

\def\beq{\begin{equation}}
\def\eeq{\end{equation}}
\def\bea{\begin{eqnarray}}
\def\eea{\end{eqnarray}}

\def\half{\frac{1}{2}}
\let\dag=\dagger

\begin{document}
\tightenlines
\thispagestyle{empty}

\font\fortssbx=cmssbx10 scaled \magstep2
\hbox to \hsize{
\hbox{\fortssbx University of Wisconsin - Madison}
\hfill
$\vcenter{\hbox{\bf AMES-HET-99-01}
\hbox{\bf ASITP-99-05}
\hbox{\bf MADPH-99-1096}
\hbox{January 1999} }$}

\vspace{.5in}

\begin{center}
{\bf NEUTRINO MIXING, $CP/T$ VIOLATION AND TEXTURES \\
IN FOUR-NEUTRINO MODELS}\\
\vskip 0.7cm
{V. Barger$^1$, Yuan-Ben Dai$^2$, K. Whisnant$^3$, and Bing-Lin Young$^3$}
\\[.1cm]
$^1${\it Department of Physics, University of Wisconsin, Madison, WI
53706, USA}
\\
$^2${\it Institute of Theoretical Physics, Chinese Academy of Sciences,
Beijing 100080, China}
\\
$^3${\it Department of Physics and Astronomy, Iowa State University,
Ames, IA 50011, USA}
\\
\end{center}

\smallskip

\begin{abstract}

We examine the prospects for determining the neutrino mixing matrix and
for observing $CP$ and $T$ violation in neutrino oscillations in
four-neutrino models. We focus on a general class of four-neutrino
models with two pairs of nearly degenerate mass eigenstates separated by
approximately 1~eV, which can describe the solar, atmospheric and LSND
neutrino data. We present a general parametrization of these models and
discuss in detail the determination of the mixing parameters and the
mass matrix texture from current and future neutrino data in the case
where $\nu_e$ and $\nu_\mu$ each mix primarily with one other
neutrino. We find that measurable $CP/T$ violation in long-baseline
experiments, with amplitude at the level of the LSND signal, is possible
given current experimental constraints. Also, additional oscillation
effects in short- and long-baseline experiments may be measurable in
many cases. We point out that, given separate scales for the
mass-squared differences of the solar and atmospheric oscillations,
observable $CP/T$ violation effects in neutrino oscillations signals the
existence of a sterile neutrino. We examine several textures of the
neutrino mass matrix and determine which textures can have measurable
$CP/T$ violation in neutrino oscillations in long-baseline
experiments. We also briefly discuss some possible origins of the
neutrino mass terms in straightforward extensions of the Standard Model.

\end{abstract}

\thispagestyle{empty}
\newpage

\section{Introduction}

Our view of the neutrino sector of the Standard Model has recently
undergone a revolutionary change. Observations of solar
neutrinos~\cite{solar1,solar2,solar3,SuperKsolar,SSM}, atmospheric
neutrinos~\cite{atmos,SuperKatmos,flux}, and accelerator
neutrinos~\cite{LSND} all indicate deviations from their predicted
values in the Standard Model with massless neutrinos.  In each case the
observation can be understood in terms of neutrino oscillations which in
turn requires nondegenerate neutrino masses.  The accelerator evidence
for oscillations is least secure, with preliminary data from the KARMEN
Collaboration \cite{KARMEN} excluding some regions of oscillation
parameters preferred by the LSND data \cite{LSND}.  Since the solar,
atmospheric, and LSND neutrino experiments have different $L/E$ (the
ratio of oscillation distance to neutrino energy), different orders of
magnitude of neutrino mass-squared differences $\delta m^2$ are
required to properly describe all features of the
data~\cite{review}. This need for three small but distinct mass-squared
differences naturally leads to the consideration of more than three
light neutrino flavors. Any additional light neutrino must be sterile,
i.e., without Standard Model gauge interactions, to be consistent with
the well-established LEP measurements of $Z \rightarrow
\nu\bar\nu$~\cite{Znunubar}. From quite general arguments it has been
shown~\cite{bgg96,bpww} that a neutrino spectrum with two pairs of
nearly degenerate mass eigenstates, separated by a gap of order 1~eV, is
required to satisfy all of the constraints from solar, atmospheric,
accelerator, and reactor data.

Sterile neutrinos (which we we denote as $\nu_s$) have long been
considered as an option for neutrino
oscillations~\cite{VB80,oldsterile}. More recently a number of models
have been proposed that utilize one or more sterile neutrinos to
describe the existing neutrino data~\cite{bpww,models,bgg98,gmnr98,roy}
or to explain r-process nucleosynthesis \cite{pelt}. However, if
sterile neutrinos mix with active flavor neutrinos they may be
stringently constrained by Big Bang nucleosynthesis (BBN). In standard
BBN phenomenology, the mass-squared difference $\delta m^2$ and the
mixing angle between a sterile and active neutrino must satisfy the
bound
\begin{equation}
\delta m^2 \sin^22\theta < 10^{-7} {\rm~eV}^2 \,,
\label{bbn}
\end{equation}
to avoid thermal overpopulation of the ``extra'', sterile neutrino
species~\cite{BBN}. The restriction in Eq.~(\ref{bbn}) would appear to
rule out all sterile-active mixing except for small-angle MSW or
vacuum mixing of solar neutrinos.  However, some recent estimates of
$N_\nu$ using higher inferred abundance of $^4$He yield a considerably
weaker bound than that given in Eq.~(\ref{bbn})~\cite{He4}. Thus BBN may
still allow sizeable mixing between sterile and active neutrinos, so
models with both small and large mixings with sterile neutrinos can be
considered.

In this paper we examine the phenomenological consequences of
four-neutrino models in which there are two pairs of neutrinos with
nearly degenerate mass eigenstates separated by about 1~eV, where the
mass separations within the pairs are several orders of magnitude
smaller.  We begin with a general parametrization of the four-neutrino
mixing matrix, and review the current experimental constraints. We then
discuss the simple situation where $\nu_e$ mixes dominantly with
$\nu_s$ or $\nu_\tau$ in solar neutrino oscillations and $\nu_\mu$ mixes
dominantly with a fourth neutrino ($\nu_\tau$ or $\nu_s$) in atmospheric
neutrino oscillations. This situation, which we refer to as the dominant
mixing scheme, has been shown to fit the existing data reasonably well.
For dominant mixing we find that the neutrino mixing matrix can be
effectively analyzed in terms of $2\times2$ blocks, where the diagonal
blocks can be approximated by simple two-neutrino rotations and the
off-diagonal blocks are small but non-vanishing. We then study the
relationship between the phenomenology of neutrino oscillations with
$CP$ violation and the texture of the neutrino mass matrix in models
where the two lightest states are much lighter than the two heaviest
states.

Since the neutrino mass matrix can in general be complex, and therefore
lead to a mixing matrix with complex elements, $CP$ violation can
naturally arise in neutrino oscillations. The pertinent question is the
size of the violation and how to observe it.  We find that if $CP$
violation exists, its size may be measurable, and has approximately the
same amplitude as indicated by the LSND experiment.  In many cases there
are small amplitude $\nu_e \rightarrow \nu_\tau$ oscillations that may
be measurable in either short- or long-baseline
experiments. Furthermore, some also have small amplitude $\nu_\mu
\rightarrow \nu_\tau$ oscillations in short-baseline experiments. We
discuss how oscillation measurements in solar, atmospheric, short-
and long-baseline neutrino experiments can, in some cases, determine
all but one of the four-neutrino mixing matrix parameters accessible to
oscillation measurements. We also discuss the minimal Higgs boson
spectrum needed to obtain the different types of four-neutrino mass
matrices, and their consequences for $CP$ violation.

This paper is organized as follows. In Sec.~II we present our
parametrization for the four-neutrino mixing matrix and expressions for
the oscillation probabilities in the case of two pairs of nearly
degenerate mass eigenstates separated by about 1~eV. We discuss the ways
in which $CP$ violation, if it exists, may be observed, and the number
of observable $CP$ violation parameters. In Sec.~III we summarize the
current constraints on the four-neutrino mixing matrix and discuss in
detail the implications of the dominant mixing scheme. We investigate
the $CP$ violation effects for several mass matrix textures in
Sec.~IV. In Sec.~V we briefly discuss some of the consequences of
neutrino mass for non-oscillation experiments, such as rare decays and
charged lepton electric dipole moments, and we emphasize the importance
in searching for these rare events, which can reveal new physics effects
other than neutrino masses. In Sec.~VI we summarize our results. Finally,
in Appendix~A we review the number of independent parameters in the
mixing matrix of Majorana neutrinos, in Appendix~B we discuss the
modest extensions of the Standard Model Higgs sector that allow us to
obtain the mass matrix textures, and in Appendix~C we determine the
neutrino mass spectrum and mixing matrix for a particular neutrino mass
matrix with $CP$ violation.

\section{Oscillation Probabilities}

\subsection{General Formalism}

We work in the basis where the charged lepton mass matrix is diagonal.
The most general neutrino mass matrix $M$ is Majorana in nature, and may
be diagonalized by a complex orthogonal transformation into a real diagonal
matrix
\beq
M_D = U^T M U \,,
\label{diag}
\eeq
by a unitary matrix $U$, which is generally obtained from the Hermitian
matrix $M^\dagger M = M^* M$ by $M_D^* M_D = U^\dagger M^\dagger M
U$\cite{mopal}. Some general properties of Majorana neutrino mass
matrices are discussed in more detail in Appendix~A. For the
four-neutrino case, labeling the flavor eigenstates by $\nu_x, \nu_e,
\nu_\mu, \nu_y$ and the mass eigenstates by $\nu_0, \nu_1, \nu_2,\nu_3$,
we may write
\beq
\left( \begin{array}{c}
\nu_x \\ \nu_e \\ \nu_\mu \\ \nu_y \\ \end{array} \right) = U
\left( \begin{array}{c}
\nu_0 \\ \nu_1 \\ \nu_2 \\ \nu_3 \\ \end{array} \right) \,.
\label{unit}
\eeq
In this paper we will examine the two cases most often considered in the
recent literature: one of $\nu_x$ and $\nu_y$ is $\nu_\tau$ and the
other is sterile ($\nu_s$), or both $\nu_x$ and $\nu_y$ are sterile.
Explicitly, the matrix $M$ may be written in the flavor basis as
\beq
M = \left( \begin{array}{cccc}
M_{xx} & M_{xe} & M_{x\mu} & M_{xy} \\
M_{xe} & M_{ee} & M_{e\mu} & M_{ey} \\
M_{x\mu} & M_{e\mu} & M_{\mu\mu} & M_{\mu y} \\
M_{xy} & M_{ey} & M_{\mu y} & M_{yy} \\ \end{array} \right) \,.
\label{Mflavor}
\eeq

The $4\times4$ unitary matrix $U$ may be parametrized by 6 rotation
angles and 6 phases, and can be conveniently represented
by~\cite{fritzsch}
\beq
U = R_{23} R_{13} R_{03} R_{12} R_{02} R_{01} \,,
\label{Urot}
\eeq
where
\beq
R_{01} = \left( \begin{array}{cccc}
c_{01} & s_{01}^* & 0 & 0 \\
-s_{01} & c_{01} & 0 & 0 \\
0 & 0 & 1 & 0 \\
0 & 0 & 0 & 1 \\
\end{array} \right) \,,
\label{rot}
\eeq
with
\beq
c_{jk} \equiv \cos\theta_{jk} \,, \qquad
s_{jk} \equiv \sin\theta_{jk} e^{i\delta_{jk}} \,,
\label{cs}
\eeq
and the other $R_{jk}$ are defined similarly for rotations in the
$j$--$k$ plane. The explicit form for the $4\times4$ unitary matrix is
\beq
U = \left( \begin{array}{cccc}
c_{01}c_{02}c_{03} & c_{02}c_{03}s_{01}^*
& c_{03}s_{02}^* & s_{03}^* \\
\\
-c_{01}c_{02}s_{03}s_{13}^*
& -c_{02}s_{01}^*s_{03}s_{13}^*
& -s_{02}^*s_{03}s_{13}^*
& c_{03}s_{13}^*
\\
-c_{01}c_{13}s_{02}s_{12}^*
& -c_{13}s_{01}^*s_{02}s_{12}^*
& +c_{02}c_{13}s_{12}^*
&
\\
-c_{12}c_{13}s_{01}
& +c_{01}c_{12}c_{13}
&
&
\\ \\
-c_{01}c_{02}c_{13}s_{03}s_{23}^*
& -c_{02}c_{13}s_{01}^*s_{03}s_{23}^*
& -c_{13}s_{02}^*s_{03}s_{23}^*
& c_{03}c_{13}s_{23}^*
\\
+c_{01}s_{02}s_{12}^*s_{13}s_{23}^*
& +s_{01}^*s_{02}s_{12}^*s_{13}s_{23}^*
& -c_{02}s_{12}^*s_{13}s_{23}^*
&
\\
-c_{01}c_{12}c_{23}s_{02}
& -c_{12}c_{23}s_{01}^*s_{02}
& +c_{02}c_{12}c_{23}
&
\\
+c_{12}s_{01}s_{13}s_{23}^*
& -c_{01}c_{12}s_{13}s_{23}^*
&
&
\\ +c_{23}s_{01}s_{12} &
-c_{01}c_{23}s_{12}
&
&
\\ \\
-c_{01}c_{02}c_{13}c_{23}s_{03}
& -c_{02}c_{13}c_{23}s_{01}^*s_{03}
& -c_{13}c_{23}s_{02}^*s_{03}
& c_{03}c_{13}c_{23}
\\
+c_{01}c_{23}s_{02}s_{12}^*s_{13}
& +c_{23}s_{01}^*s_{02}s_{12}^*s_{13}
& -c_{02}c_{23}s_{12}^*s_{13}
&
\\
+c_{01}c_{12}s_{02}s_{23}
& +c_{12}s_{01}^*s_{02}s_{23}
& -c_{02}c_{12}s_{23}
&
\\
+c_{12}c_{23}s_{01}s_{13}
& -c_{01}c_{12}c_{23}s_{13}
&
&
\\
-s_{01}s_{12}s_{23}
& +c_{01}s_{12}s_{23}
&
&
\\ \\ \end{array} \right) \,.
\label{genU}
\eeq
We will label the matrix elements of $U$ by $U_{\alpha j}$, where Greek
indices denote flavor eigenstate labels ($\alpha=x,e,\mu,y$) and Latin
indices denote mass eigenstate labels ($j=0,1,2,3$). With the knowledge
of the mass eigenvalues and mixing matrix elements, one can invert
Eq.~(\ref{diag}) to obtain the neutrino mass matrix elements
\beq
M_{\alpha\beta} = \sum_{j=0}^3 U_{\alpha j}^* U_{\beta j}^* m_j \,.
\label{Mab}
\eeq

The vacuum neutrino flavor oscillation probabilities, for an initially
produced $\nu_\alpha$ to a finally detected $\nu_\beta$, can be written
\beq
P(\nu_\alpha \rightarrow \nu_\beta) =
\delta_{\alpha\beta} - \sum_{j<k} \left[
4 {\rm~Re}(W^{jk}_{\alpha\beta}) \sin^2\Delta_{kj}
- 2 {\rm~Im}(W^{jk}_{\alpha\beta}) \sin2\Delta_{kj} \right] \,,
\label{genprob}
\eeq
where
\bea
W^{jk}_{\alpha\beta} &\equiv
&
U_{\alpha j} U_{\alpha k}^* U_{\beta j}^* U_{\beta k} \,,
\label{W}
\\
\Delta_{kj} &\equiv& \delta m^2_{kj} L/(4E) \,, \qquad
\delta m^2_{kj} \equiv m_k^2 - m_j^2 \,,
\label{Delta}
\eea
$L$ is the oscillation distance, and $E$ is the neutrino energy.
The quantities $W^{jk}_{\alpha\beta}$ \cite{Weiler}, are related to the
Jarlskog invariants~\cite{jarlskog}
\beq
J^{jk}_{\alpha\beta} \equiv {\rm~Im}(W^{jk}_{\alpha\beta}) \,,
\label{J}
\eeq
which satisfy the identity
\beq
J^{jk}_{\alpha\beta} = - J^{jk*}_{\alpha\beta} \,,
\label{Jcc}
\eeq
obtained by the interchange $U \leftrightarrow U^*$ in Eq.~(\ref{W}).
Also,
\beq
J^{jk}_{\alpha\beta} = J^{kj}_{\beta\alpha} =
- J^{jk}_{\beta\alpha} = - J^{kj}_{\alpha\beta} \,.
\label{Jasy}
\eeq
We can also define the real part of $W^{jk}_{\alpha\beta}$ as
\beq
Y^{jk}_{\alpha\beta} \equiv {\rm Re}(W^{jk}_{\alpha\beta}) \,,
\label{Y}
\eeq
which is invariant under interchange of the upper or lower indices:
\beq
Y^{jk}_{\alpha\beta} = Y^{kj}_{\beta\alpha} =
Y^{jk}_{\beta\alpha} = Y^{kj}_{\alpha\beta} \,.
\label{Ysym}
\eeq
Another useful property of the $W^{jk}_{\alpha\beta}$ is that the sum
over any of the indices reduces them to a real positive quantity, e.g.,
\bea
\sum_\beta W^{jk}_{\alpha\beta} &=& |U_{\alpha j}|^2 \delta_{jk}
= \sum_\beta Y^{jk}_{\alpha\beta} \,,
\label{Ysum}
\\
\sum_\beta J^{jk}_{\alpha\beta} &=& 0 \,.
\label{Jsum}
\eea
Equations~(\ref{genprob}), (\ref{Jcc}), (\ref{Jasy}) and (\ref{Ysym})
imply
\beq
P(\nu_\alpha \rightarrow \nu_\beta) =
P(\bar\nu_\beta \rightarrow \bar\nu_\alpha) \,,
\label{JCPT}
\eeq
which is a statement of $CPT$ invariance. Equation~(\ref{genprob}) and
(\ref{Jasy}) imply
that nonzero $J^{jk}_{\alpha\beta}$ can give $CP$ or $T$ violation
\beq
P(\nu_\alpha \rightarrow \nu_\beta) \ne
P(\bar\nu_\alpha \rightarrow \bar\nu_\beta) =
P(\nu_\beta \rightarrow \nu_\alpha) \,.
\label{P}
\eeq
{}From Eq.~(\ref{P}) we can define the $CP$-violation quantity
\beq
\Delta P_{\alpha\beta} =
P(\nu_\alpha \rightarrow \nu_\beta) -
P(\nu_\beta \rightarrow \nu_\alpha) \,.
\label{deltap}
\eeq
In four-neutrino oscillations there are only three
independent $\Delta P_{\alpha\beta}$, and, correspondingly, only three
of the six phases in $U$ can be measured in neutrino oscillations (for a
discussion, see Appendix~A). Thus six angles and three phases can in
principle be measured in neutrino oscillations, which is the same as in
the Dirac neutrino case. Therefore, as far as neutrino oscillations are
concerned, our results apply equally to Dirac neutrinos. The three
remaining independent phases in $U$ enter into the mass matrix
elements and processes such as neutrinoless double beta decay.

\subsection{Model with two nearly degenerate pairs of neutrinos}

For a four-neutrino model to describe the solar, atmospheric, LSND data
and also satisfy all other accelerator and reactor limits, it must have
two pairs of nearly degenerate mass eigenstates~\cite{bgg96,bpww}; e.g.,
$\delta m^2_{sun} \equiv \delta m^2_{01} \ll \delta m^2_{atm} \equiv
\delta m^2_{32} \ll \delta m^2_{LSND} \equiv \delta m^2_{21}$. We will
also assume without loss of generality that $0 < m_0, m_1 < m_2 < m_3$.
An alternative scenario with the roles of $\delta m^2_{01}$ and $\delta
m^2_{32}$ reversed gives the same results as far as oscillations are
concerned, although the implications for the mass matrix, double beta
decay and cosmology may differ; this alternate possibility will be
briefly discussed in Sec.~IV.E. Also note that if the solar oscillations
are driven by the MSW effect~\cite{MSW}, we must require $m_0 > m_1$;
for vacuum oscillations, $m_0 < m_1$ is also possible.

Given this hierarchy of the $\delta m^2$, the oscillation probabilities
for $\alpha\ne\beta$ may be written approximately as
\bea
P(\nu_\alpha \rightarrow \nu_\beta) &\simeq&
A^{\alpha\beta}_{LSND} \sin^2\Delta_{LSND}
+A^{\alpha\beta}_{atm} \sin^2\Delta_{atm}
+B^{\alpha\beta}_{atm} \sin2\Delta_{atm}
\nonumber \\
&&+A^{\alpha\beta}_{sun} \sin^2\Delta_{sun}
+B^{\alpha\beta}_{sun} \sin2\Delta_{sun} \,, \qquad \alpha \ne \beta \,,
\label{offprob}
\eea
and for the diagonal channels
\beq
P(\nu_\alpha \rightarrow \nu_\alpha) \simeq
1 - A^{\alpha\alpha}_{LSND} \sin^2\Delta_{LSND}
- A^{\alpha\alpha}_{atm} \sin^2\Delta_{atm}
- A^{\alpha\alpha}_{sun} \sin^2\Delta_{sun} \,,
\label{diagprob}
\eeq
where $\Delta_{scale} \equiv {1\over4} \delta m^2_{scale} L/E$,
$A^{\alpha\beta}_{scale}$ is the usual $CP$ conserving oscillation
amplitude for $\nu_\alpha \rightarrow \nu_\beta$ at a given oscillation
scale, $B^{\alpha\beta}_{scale}$ is the $CP$ violation parameter at a
given scale, and the scale label is $sun$ for the solar neutrino scale,
$atm$ for atmospheric and long-baseline scales, and {\small$LSND$} for
accelerator and short-baseline scales.
Note that the $CP$-violating terms have a different dependence on $L/E$
from the $CP$-conserving terms, which could in principle be distinguished by
measurements at different $L/E$~\cite{pakvasa}.
In Eqs.~(\ref{offprob}) and
(\ref{diagprob}) we have used the approximation $\Delta_{31} \simeq
\Delta_{30} \simeq \Delta_{21} \simeq \Delta_{20} \simeq \Delta_{LSND}$.

The oscillation amplitudes are given by
\bea
A^{\alpha\beta}_{LSND} &=&
4 |U_{\alpha 2} U_{\beta 2}^* + U_{\alpha 3} U_{\beta 3}^*|^2
= 4 |U_{\alpha 0} U_{\beta 0}^* + U_{\alpha 1} U_{\beta 1}^*|^2
\,, \qquad \alpha\ne\beta \,,
\label{Asbloff}
\\
A^{\alpha\alpha}_{LSND}
&=& 4 (|U_{\alpha 2}|^2 + |U_{\alpha 3}|^2)
(1 - |U_{\alpha 2}|^2 - |U_{\alpha 3}|^2) \,,
\nonumber \\
&=& 4 (|U_{\alpha 0}|^2 + |U_{\alpha 1}|^2)
(1 - |U_{\alpha 0}|^2 - |U_{\alpha 1}|^2) \,,
\label{Asbldiag}
\\
A^{\alpha\beta}_{atm} &=&
- 4 {\rm~Re}(U_{\alpha 2} U_{\alpha 3}^* U_{\beta 2}^* U_{\beta 3})
\,, \qquad \alpha\ne\beta \,,
\label{Aatmoff}
\\
A^{\alpha\alpha}_{atm} &=&
4|U_{\alpha 2}|^2 |U_{\alpha 3}|^2 \,,
\label{Aatmdiag}
\\
A^{\alpha\beta}_{sun} &=&
- 4 {\rm~Re}(U_{\alpha 0} U_{\alpha 1}^* U_{\beta 0}^* U_{\beta 1})
\,, \qquad \alpha\ne\beta \,,
\label{Asunoff}
\\
A^{\alpha\alpha}_{sun} &=&
4 |U_{\alpha 0}|^2 |U_{\alpha 1}|^2 \,,
\label{Asundiag}
\eea
where the second equality in Eqs.~(\ref{Asbloff}) and (\ref{Asbldiag})
follows from the unitarity of $U$. We note that the form of the
short-baseline oscillation amplitudes in Eqs.~(\ref{Asbloff}) and
(\ref{Asbldiag}) are different from the cases of long-baseline,
Eqs.~(\ref{Aatmoff}) and (\ref{Aatmdiag}), and solar,
Eqs.~(\ref{Asunoff}) and (\ref{Asundiag}). The difference is due
to the fact that the short-baseline oscillations arise from four
mass-squared differences ($\delta m^2_{20} \simeq \delta m^2_{30}
\simeq \delta m^2_{21} \simeq \delta m^2_{31}$), while the
long-baseline and solar oscillations arise from only one mass-squared
difference ($\delta m^2_{32}$ and $\delta m^2_{01}$,
respectively). Probability conservation implies
$A^{\alpha\alpha}_{scale} = \sum_{\beta\ne\alpha}
A^{\alpha\beta}_{scale}$, which can easily be shown using the unitarity
of $U$. The $CP$ violation parameters are
\bea
B^{\alpha\beta}_{atm} &=&
- 2 {\rm~Im}(U_{\alpha 2} U_{\alpha 3}^* U_{\beta 2}^* U_{\beta 3}) \,,
\label{Batmoff}
\\
B^{\alpha\beta}_{sun} &=&
2 {\rm~Im}(U_{\alpha 0} U_{\alpha 1}^* U_{\beta 0}^* U_{\beta 1}) \,.
\label{Bsunoff}
\eea
Since $B^{\alpha\alpha}_j = 0$, there is no $CP$ violation in diagonal
channels.  The absence of $B^{\alpha\beta}_{LSND}$ in Eq.~(\ref{offprob})
shows that no observable $CP$ violation is present for the leading
oscillation~\cite{CP3}, and $CP$ violation may only be seen in experiments
that probe non-leading scales, $\delta m^2_{atm}$ or $\delta m^2_{sun}$.

For short-baseline experiments where only the leading oscillation
argument $\Delta_{LSND}$ has had a chance to develop, the off-diagonal
vacuum oscillation probabilities are
\bea
P(\nu_\alpha \rightarrow \nu_\beta) &\simeq&
A^{\alpha\beta}_{LSND} \sin^2\Delta_{LSND} \,, \qquad \alpha\ne\beta \,,
\label{sbloff}
\\
P(\nu_\alpha \rightarrow \nu_\alpha) &\simeq&
1 - A^{\alpha\alpha}_{LSND} \sin^2\Delta_{LSND} \,.
\label{sbldiag}
\eea
For larger $L/E$ (such as in atmospheric and long-baseline
experiments), where the secondary oscillation has had time to
develop, the vacuum oscillation probabilities are
\bea
P(\nu_\alpha \rightarrow \nu_\beta) &\simeq&
{1\over2} A^{\alpha\beta}_{LSND}
+ A^{\alpha\beta}_{atm}\sin^2\Delta_{atm}
+ B^{\alpha\beta}_{atm}\sin2\Delta_{atm} \,, \qquad \alpha\ne\beta \,,
\label{atmoff}
\\
P(\nu_\alpha \rightarrow \nu_\alpha) &\simeq&
1 - {1\over2}A^{\alpha\alpha}_{LSND}
- A^{\alpha\alpha}_{atm} \sin^2\Delta_{atm} \,.
\label{atmdiag}
\eea
Here we have assumed that the leading oscillation has averaged, i.e.,
$\sin^2\Delta_{LSND} \rightarrow {1\over2}$. Finally, at the solar
distance scale, when all oscillation effects have developed, the
vacuum oscillation probabilities are
\bea
P(\nu_\alpha \rightarrow \nu_\beta) &\simeq&
{1\over2}(A^{\alpha\beta}_{LSND}+A^{\alpha\beta}_{atm})
+ A^{\alpha\beta}_{sun} \sin^2\Delta_{sun}
+ B^{\alpha\beta}_{sun} \sin2\Delta_{sun} \,, \qquad \alpha\ne\beta \,,
\label{sunoff}
\\
P(\nu_\alpha \rightarrow \nu_\alpha) &\simeq&
1 - {1\over2}(A^{\alpha\alpha}_{LSND}+A^{\alpha\alpha}_{atm})
- A^{\alpha\alpha}_{sun} \sin^2\Delta_{sun} \,,
\label{sundiag}
\eea
where $\sin^2\Delta_{atm}$ has been averaged to ${1\over2}$ and
$\sin2\Delta_{atm}$ has been averaged to $0$; the $CP$ violation at the
$\delta m^2_{atm}$ scale is washed out. $CP$ violation is possible only
in the off-diagonal channels, as noted before, and the solar neutrino
$\nu_e$ survival measurement cannot be used to observed $CP$ violation.

More generally, in any model for which the oscillation scales are
well-separated and $L/E$ is only large enough to probe the largest
oscillation scale, $CP$-violating effects in neutrino oscillations will
be unobservable (strictly speaking, they are suppressed to order $\Delta
\ll 1$, where $\Delta$ is the oscillation argument for the
second-largest oscillation scale)~\cite{CP3}. The $CP$-violating
effects become observable when $L/E$ is large enough to probe both the
largest and second-largest oscillation scales. For a four-neutrino
model with different oscillation scales to describe the solar,
atmospheric and LSND data, this means that $CP$ violation can only be
detected in experiments with $L/E$ at least as large as those found in
atmospheric and long-baseline experiments.

As a corollary, in three-neutrino models with two oscillation scales
describing only the solar and atmospheric data, $CP$ violation has the
potential to be observable only in experiments with $L/E$ comparable to
or larger than the solar experiments. However, a measurement of
off-diagonal oscillation probabilities is required to see $CP$
violation, and that is not possible in solar neutrino experiments.
Hence, if the solar neutrino oscillation scale is well-established, the
observation of a $CP$-violation effect in long-baseline experiments
could imply that there are at least three separate neutrino
mass-squared difference scales, and thus more than three neutrinos.

\section{Determining the oscillation parameters}

In this section we first derive some general constraints imposed by
current data on the neutrino mixing matrix for four-neutrino models
favored by the data, i.e., with two pairs of nearly degenerate masses
satisfying $\delta m^2_{01} \ll \delta m^2_{32} \ll \delta m^2_{21}$. We
then determine the form of the mixing matrix under the assumption that
$\nu_e$ and $\nu_x$ are mostly a mixture of $\nu_0$ and $\nu_1$ and
provide the dominant solar neutrino oscillation, and $\nu_\mu$ and
$\nu_y$ are mostly a mixture of $\nu_2$ and $\nu_3$ and provide the
dominant atmospheric neutrino oscillation, which is the form of most
explicit models in the literature. Then we discuss the measurements
needed to determine the parameters in the neutrino mixing matrix. The
results of this section apply equally to the case where $m_2 < m_3 <
m_0,m_1$. The more general case where $\nu_e$ and $\nu_\mu$ have large
mixing with more than one other neutrino is briefly discussed in
Sec.~IV.E.

\subsection{Solar $\nu_e \rightarrow \nu_x$ and atmospheric $\nu_\mu
\rightarrow \nu_y$}

The flavor eigenstates are related to the mass eigenstates by
Eq.~(\ref{unit}). Using the formulae in Sec.~II, the amplitudes
for short-baseline oscillation (such as LSND, reactors, and
other past accelerator oscillation searches) are
\bea
A^{e\mu}_{LSND} &=& 4|U_{e2}U_{\mu2}^* + U_{e3}U_{\mu3}^*|^2
= 4|U_{e0}U_{\mu0}^* + U_{e1}U_{\mu1}^*|^2 \,,
\label{Aemsbl}
\\
A^{\mu\mu}_{LSND}
&=& 4(|U_{\mu2}|^2+|U_{\mu3}|^2)(1-|U_{\mu2}|^2-|U_{\mu3}|^2)
\nonumber \\
&=& 4(|U_{\mu0}|^2+|U_{\mu1}|^2)(1-|U_{\mu0}|^2-|U_{\mu1}|^2) \,,
\label{Ammsbl}
\\
A^{ee}_{LSND}
&=& 4(|U_{e2}|^2+|U_{e3}|^2)(1-|U_{e2}|^2-|U_{e3}|^2)
\nonumber \\
&=& 4(|U_{e0}|^2+|U_{e1}|^2)(1-|U_{e0}|^2-|U_{e1}|^2) \,,
\label{Aeesbl}
\eea
where the second equalities in each case result from the unitarity of
$U$. For atmospheric and long-baseline oscillation, the amplitudes are
\bea
A^{\mu\mu}_{atm} &=& 4|U_{\mu2}|^2|U_{\mu3}|^2 \,,
\label{Ammatm}
\\
A^{e\mu}_{atm} &=& -4{\rm~Re}(U_{e2}U_{e3}^*U_{\mu2}^*U_{\mu3}) \,,
\label{Aematm}
\\
B^{e\mu}_{atm} &=& -2{\rm~Im}(U_{e2}U_{e3}^*U_{\mu2}^*U_{\mu3}) \,,
\label{Bematm}
\\
A^{ey}_{atm} &=& -4{\rm~Re}(U_{e2}U_{e3}^*U_{y2}^*U_{y3}) \,,
\label{Aeyatm}
\\
B^{ey}_{atm} &=& -2{\rm~Im}(U_{e2}U_{e3}^*U_{y2}^*U_{y3}) \,,
\label{Beyatm}
\\
A^{\mu y}_{atm} &=& -4{\rm~Re}(U_{\mu2}U_{\mu3}^*U_{y2}^*U_{y3}) \,,
\label{Amyatm}
\\
B^{\mu y}_{atm} &=& -2{\rm~Im}(U_{\mu2}U_{\mu3}^*U_{y2}^*U_{y3}) \,,
\label{Bmyatm}
\eea
and in solar experiments
\beq
A^{ee}_{sun} = 4 |U_{e0}|^2 |U_{e1}|^2 \,.
\label{Aeesun}
\eeq

The atmospheric neutrino experiments favor large mixing of $\nu_\mu$
at the atmospheric scale \cite{bpww,fogli}
\beq
A^{\mu\mu}_{atm} > 0.8 \,,
\eeq
at 90\%~C.L.; then Eq.~(\ref{Ammatm}) and unitarity imply
\beq
|U_{\mu2}|^2 + |U_{\mu3}|^2 > 0.894 \,, \qquad
|U_{\mu0}|^2 + |U_{\mu1}|^2 < 0.106 \,.
\label{maxmix}
\eeq
Also, the Bugey reactor constraint~\cite{bugey} gives
\beq
A^{ee}_{LSND} < 0.06 \,,
\eeq
over the indicated range for $\delta m^2_{LSND}$; then
Eq.~(\ref{Aeesbl}) implies
\beq
|U_{e2}|^2 + |U_{e3}|^2 < 0.016 \,, \qquad
|U_{e0}|^2 + |U_{e1}|^2 > 0.984 \,.
\label{bugey}
\eeq
Finally, the oscillation interpretation of the LSND results~\cite{LSND}
gives
\beq
A^{e\mu}_{LSND} \equiv \eps^2 \,,
\eeq
where $\eps$ is experimentally constrained to the range~\cite{LSND}
\beq
0.05 < \eps < 0.20 \,,
\label{eps}
\eeq
where the exact value depends on $\delta m^2_{LSND}$. Hence, the atmospheric
and Bugey results imply that $|U_{e2}|$, $|U_{e3}|$, $|U_{\mu0}|$, and
$|U_{\mu1}|$ are all approximately of order $\eps$ or smaller. We note
that given these constraints the size of $A^{\mu\mu}_{LSND}$ must also
be small, in agreement with the CDHS bound on $\nu_\mu$
disappearance~\cite{CDHS}.

If we assume that it is only $\nu_y$ that mixes appreciably with
$\nu_\mu$ in the atmospheric experiments and only $\nu_x$ that mixes
appreciably with $\nu_e$ in the solar experiments, then $|U_{x2}|$,
$|U_{x3}|$, $|U_{y0}|$, and $|U_{y1}|$ must also be small, i.e., of
order $\eps$ or less.  The mixing matrix can therefore be seen to have
the form
\beq
U
=\left( \begin{array}{c|c}
U_1 & U_2 \\
\hline
U_3 & U_4 \\
\end{array} \right) \,,
\label{Ueps}
\eeq
where the $U_j$ ($j=1,2,3,4$) are $2\times2$ matrices and the elements
of $U_2$ and $U_3$ are at most of order $\eps$ in size. The matrix $U_4$
is approximately the $2\times2$ maximal mixing matrix (i.e., all elements
have approximate magnitude $1/\sqrt2$) that describes atmospheric $\nu_\mu
\rightarrow \nu_y$ oscillations, but $U_1$, which is approximately
unitary by itself and which primarily  describes the mixing in the
solar neutrino sector, may have large (for vacuum oscillations) or small
(for MSW oscillations) mixing.

The form of $U$ in Eq.~(\ref{Ueps}), with $U_2$ and $U_3 \sim \eps$,
implies
\beq
|s_{02}|, |s_{03}|, |s_{12}|, |s_{13}| \sim \eps \,,
\label{seps}
\eeq
in the general parametrization of Eq.~(\ref{genU}). After
dropping terms second order in $\eps$ and smaller, $U$ takes the form
\beq
U \simeq \left( \begin{array}{cccc}
c_{01} & s_{01}^* & s_{02}^* & s_{03}^*
\\
&&&\\
-s_{01} & c_{01} & s_{12}^* & s_{13}^*
\\
&&&\\
-c_{01}(s_{23}^*s_{03}+c_{23}s_{02})
& -s_{01}^*(s_{23}^*s_{03}+c_{23}s_{02})
& c_{23}
& s_{23}^*
\\
+s_{01}(s_{23}^*s_{13}+c_{23}s_{12})
& -c_{01}(s_{23}^*s_{13}+c_{23}s_{12})
&&\\
&&&\\
c_{01}(s_{23}s_{02}-c_{23}s_{03})
& s_{01}^*(s_{23}s_{02}-c_{23}s_{03})
& -s_{23}
& c_{23}
\\
-s_{01}(s_{23}s_{12}-c_{23}s_{13})
& +c_{01}(s_{23}s_{12}-c_{23}s_{13})
&&\\
&&&\\
\end{array} \right) \,.
\label{genU2}
\eeq
This matrix provides a general parametrization of the
four-neutrino mixing in models where $\nu_e$ mixes primarily with
$\nu_x$ at the solar mass-squared difference scale, and $\nu_\mu$ mixes
primarily with $\nu_y$ at the atmospheric mass-squared difference
scale. Unitarity of $U$ is satisfied to the order of $\eps$.

Despite the fact that the expansion of the matrix elements of $U$ in
Eq.~(\ref{genU2}) is to the first order of $\eps$, it still allows us,
as shown below, to extract all of the interesting oscillation and $CP$
violation effects, which are second order in $\eps$. Care must be
taken when the leading order result cancels, and sometimes it is helpful
to use the unitarity of $U$ to derive an alternate expression that gives
the correct leading order answer, e.g., for $A^{\mu y}_{LSND}$, the
first expression in Eq.~(\ref{Asbloff}) gives zero when the form of $U$
in Eq.~(\ref{genU2}) is used, but the second expression gives a finite
(and correct to leading order) result.

The off-diagonal oscillation amplitudes for the leading oscillation
are
\bea
A^{e\mu}_{LSND} &=& 4|s_{12}c_{23} + s_{13}s_{23}^*|^2 \,,
\label{Aemsbl2}
\\
A^{ey}_{LSND} &=& 4|s_{12}s_{23} - s_{13}c_{23}|^2 \,,
\label{Aeysbl2}
\\
A^{\mu x}_{LSND} &=& 4|s_{02}c_{23} + s_{03}s_{23}^*|^2 \,,
\label{Amxsbl2}
\\
A^{ex}_{LSND}  = A^{\mu y}_{LSND} &=& O(\eps^4) \,.
\label{Aexsbl2}
\eea
For atmospheric and long-baseline experiments the oscillation
amplitudes are
\bea
A^{e\mu}_{atm} = - A^{ey}_{atm} &=&
-4 c_{23}{\rm Re}(s_{12}^*s_{13}s_{23}^*) \,,
\label{Aematm2}
\\
B^{e\mu}_{atm} = - B^{ey}_{atm} &=&
-2 c_{23}{\rm Im}(s_{12}^*s_{13}s_{23}^*) \,,
\label{Bematm2}
\\
A^{ex}_{atm}, B^{ex}_{atm} &=& O(\eps^4) \,,
\label{Aexatm2}
\\
A^{\mu\mu}_{atm} \simeq A^{\mu y}_{atm} &=& \sin^22\theta_{23} \,,
\label{Ammatm2}
\\
A^{\mu x}_{atm} &=&
-4 c_{23}{\rm Re}(s_{02}^*s_{03}s_{23}^*) \,,
\label{Amxatm2}
\\
B^{\mu x}_{atm} &=&
2 c_{23}{\rm Im}(s_{02}^*s_{03}s_{23}^*) \,,
\label{Bmxatm2}
\\
B^{\mu y}_{atm} &=&
-2 c_{23} {\rm Im}[(s_{02}^*s_{03}+s_{12}^*s_{13})s_{23}^*] \,,
\label{Bmyatm2}
\eea
where $\theta_{23}$ is defined in Eq.~(\ref{cs}). All
oscillation amplitudes at the LSND scale and all oscillation amplitudes
(including the $CP$ violation amplitudes) other than $A^{\mu y}_{atm}$
at the atmospheric scale are at most of order $\eps^2$. Since $A^{\mu
y}_{atm}$ is large, $B^{\mu y}_{atm}$, which is of order $\eps^2$, may
be hard to measure since it involves taking the difference of two nearly
equal large numbers. At the solar scale, we have
\beq
A^{ee}_{sun} \simeq A^{ex}_{sun} = \sin^22\theta_{01} \,.
\label{Asun}
\eeq

There are 6 mixing angles and 3 independent phases that in principle may
be measured in neutrino oscillations. In all, there are eight
independent parameters involved in the
observables in Eqs.~(\ref{Aemsbl2})--(\ref{Asun}), which are the six
mixing angles $\theta_{01}$, $\theta_{02}$, $\theta_{03}$,
$\theta_{12}$, $\theta_{13}$, $\theta_{23}$, and the two phases
\bea
\phi_0 &\equiv& \delta_{03} - \delta_{02} - \delta_{23} \,,
\label{phi0}
\\
\phi_1 &\equiv& \delta_{13} - \delta_{12} - \delta_{23} \,.
\label{phi1}
\eea
These eight parameters could in principle be determined by measurements
of the eight observables $A^{ee}_{sun}$, $A^{\mu\mu}_{atm}$,
$A^{e\mu}_{LSND}$, $A^{e\mu}_{atm}$, $B^{e\mu}_{atm}$, $A^{\mu
x}_{LSND}$, $A^{\mu x}_{atm}$, and $B^{\mu x}_{atm}$. Therefore if
$\nu_x = \nu_\tau$, then all eight of these parameters could in
principle be determined from the solar, atmospheric, short- and
long-baseline experiments. This emphasizes the need for both short-
and long-baseline measurements of all active oscillation channels,
since the oscillation amplitudes involve different combinations of the
parameters at short and long baselines. If $\nu_x$ is sterile, the three
parameters $\theta_{02}$, $\theta_{03}$ and $\phi_0$ might be difficult
to determine since they would involve the disappearance $\nu_\mu
\rightarrow \nu_s$ that is at most of order $\eps^2$ in magnitude. If
$\nu_y = \nu_\tau$, the additional observables $A^{e y}_{LSND}$,
$A^{ey}_{atm}$, and $B^{ey}_{atm}$ can provide a consistency check on
the parameters $\theta_{12}$, $\theta_{13}$, and $\phi_1$.

We note that from the above results that many of the $CP$-violating
amplitudes can be the same order of magnitude as the corresponding
$CP$-conserving amplitudes, and hence potentially observable in
high-statistics long-baseline experiments. The $CP$ violation
parameters $B^{e\mu}_{atm}$ and $B^{\mu x}_{atm}$ could be determined in
vacuum by measuring probability differences $\Delta \bar P_{e\mu}$ and
$\Delta \bar P_{\mu x}$, where
\beq
\Delta \bar P_{\alpha\beta} \equiv
P(\nu_\alpha \rightarrow \nu_\beta) - P(\bar\nu_\alpha \rightarrow
\bar\nu_\beta) \,,
\label{delPbar}
\eeq
in long-baseline experiments, or the probability differences $\Delta
P_{e\mu}$ and $\Delta P_{\mu x}$, where $\Delta P_{\alpha\beta}$,
defined in Eq.~(\ref{deltap}), measures explicit $T$-violation. In a
vacuum,
\beq
\Delta P_{\alpha\beta} =
\Delta \bar P_{\alpha\beta} = 2 B^{\alpha\beta}_{atm} \sin2\Delta_{atm}
\,.
\eeq
Alternatively, one could measure $CP$ asymmetries
\beq
{\cal A}^{CP}_{\alpha\beta} = { P(\nu_\alpha \rightarrow \nu_\beta) -
P(\bar\nu_\alpha \rightarrow \bar\nu_\beta) \over
P(\nu_\alpha \rightarrow \nu_\beta) +
P(\bar\nu_\alpha \rightarrow \bar\nu_\beta)} \,,
\label{ACP}
\eeq
or $T$ asymmetries
\beq
{\cal A}^T_{\alpha\beta}
= { P(\nu_\alpha \rightarrow \nu_\beta) -
P(\nu_\beta \rightarrow \nu_\alpha) \over
P(\nu_\alpha \rightarrow \nu_\beta) +
P(\nu_\beta \rightarrow \nu_\alpha)} \,.
\label{AT}
\eeq
In vacuum, $CPT$ invariance insures that $\Delta P_{\alpha\beta} =
\Delta \bar P_{\alpha\beta}$ and ${\cal A}^T_{\alpha\beta} = {\cal
A}^{CP}_{\alpha\beta}$. However, matter effects could induce a nonzero
$\Delta \bar P_{\alpha\beta}$ or ${\cal A}^{CP}_{\alpha\beta}$ even in
the absence of $CP$ violation~\cite{bgg98,MN}. Since the matter
effects in long-baseline experiments for $P(\nu_\alpha \rightarrow
\nu_\beta)$ and $P(\nu_\beta \rightarrow \nu_\alpha)$ are the same, the
quantities $\Delta P_{e\mu}$ and ${\cal A}^T_{e\mu}$, which can only be
nonzero if there is explicit $CP$ or $T$ violation, may be
preferable~\cite{bgg98}.

There remains a third independent phase that could have consequences for
neutrino oscillations, but will in practice be difficult to
measure. This phase, which could be taken as $\delta_{01}$ defined in
Eq.~(\ref{Urot}), could be determined from $CP$ violation in $\nu_e
\leftrightarrow \nu_\mu$ or $\nu_y$ at the $\delta m^2_{sun}$ scale, but
this effect would require the
measurement of an off-diagonal channel at the solar scale. Therefore it
appears that a complete determination of the four-neutrino mixing
matrix is not possible with conventional oscillation experiments.
Table~\ref{params} lists all parameters that appear to be accessible to
observation together with the principal observables that determine these
parameters.

\subsection{More general mixing scenarios}

In general, both solar $\nu_e$ and atmospheric $\nu_\mu$ could oscillate
into mixtures of $\nu_x$ and $\nu_y$. In this event $\theta_{02}$ and
$\theta_{03}$ in Eq.~(\ref{genU}) are not necessarily small. If one of
$\nu_x$ and $\nu_y$ is the tau neutrino and the other sterile, there are
several possible ways that the existence of such  mixing could be
determined~\cite{bpww2}. Also, vacuum $CP$-violation effects involving
$\nu_e$ will still be no larger than order $\eps^2$ (due to the
smallness of $U_{e2}$ and $U_{e3}$), but there are potentially large
$CP$-violation effects in long-baseline $\nu_\mu$-$\nu_y$
oscillations (as large as allowed by the unitarity of $U$)~\cite{bgg98}.

\section{$CP$ violation and neutrino mass textures}

In this section we study the relationship between the neutrino mass
texture and the possibility for observable $CP$ violation (and,
equivalently, $T$ violation) in neutrino oscillations in four-neutrino
models. We will consider models where one of $\nu_x$ and $\nu_y$ is
sterile and the other is $\nu_\tau$ (such as in Refs.~\cite{bpww},
\cite{gmnr98}, \cite{roy}, and \cite{mohap}), and also models where
both are sterile~\cite{krolikowski}, which are two possible extensions
of the Standard Model neutrinos. Note that in all earlier studies the
mass matrices were taken to be real and no $CP$ violation was possible.
In Appendix~B we discuss straightforward extensions of the Standard
Model for the two cases and show explicitly how their neutrino mass
matrices can arise. In Secs.~IV.A--IV.D we assume that
\beq
m_0 \simeq m_1 \ll m_2 \simeq m_3 \simeq \sqrt{\delta m^2_{LSND}} \,,
\eeq
i.e., the lighter pair of nearly-degenerate mass eigenstates are much
lighter than the heavier pair, also nearly degenerate, which is the
structure of most explicit four-neutrino models in the literature.
In Sec.~IV.E we briefly discuss models with other mass hierarchies.

{}From Eqs.~(\ref{Mab}) and (\ref{genU2}) the neutrino mass matrix
elements involving $\nu_\mu$ and $\nu_y$ are, to leading order in $\eps$,
\bea
M_{\mu\mu} \simeq M_{yy}^* &\simeq& m (c_{23}^2 + s_{23}^2) \,,
\label{Mmm2}
\\
M_{\mu y} &\simeq& i m \sin2\theta_{23}\sin\delta_{23} \,,
\label{Mmy2}
\\
M_{e\mu} &\simeq& m (s_{12}c_{23} + s_{13}s_{23}) \,,
\label{Mem2}
\\
M_{ey} &\simeq& m (s_{13}c_{23} - s_{12}s_{23}^*) \,,
\label{Mey2}
\\
M_{x\mu} &\simeq& m (s_{02}c_{23} + s_{03}s_{23}) \,,
\label{Mxm2}
\\
M_{xy} &\simeq& m (s_{03}c_{23} - s_{02}s_{23}^*) \,,
\label{Mxy2}
\eea
where we have used the relative sizes of the $U_{\alpha j}$ and mass
eigenvalues, and the fact that to leading order in $\eps$, $m_2 \simeq
m_3 \equiv m$.  Note that the $s_{jk}$, defined in Eq.~(\ref{cs}), may
be complex. For the mass matrix elements $M_{xx}$, $M_{xe}$, and
$M_{ee}$, all four terms in Eq.~(\ref{Mab}) are small and may be of
similar size (the first two are suppressed by the small values of $m_0$
and $m_1$, the last two by mixing angles of size $\eps$); their values
depend on the exact structure in the solar sector, which we do not
specify here.  More precise solar neutrino measurements would help to
determine their values.

The three phases which enter in the expressions for the mass matrix
elements given above are $\delta_{23}$ and
\bea
\phi_0^\prime \equiv \delta_{03} - \delta_{02} + \delta_{23} \,,
\label{phi0prime}
\\
\phi_1^\prime \equiv \delta_{13} - \delta_{12} + \delta_{23} \,.
\label{phi1prime}
\eea
Only the phases $\phi_0$ and $\phi_1$, which can be measured in
oscillation experiments, and $\phi_0^\prime$, which appears in the
expressions for the mass matrix elements, are independent. The two
phases $\delta_{23} = (\phi_0^\prime - \phi_0)/2$ and $\phi_1^\prime =
\phi_0^\prime +\phi_1 - \phi_0$ are linearly dependent.

Equations~(\ref{Mmm2})--(\ref{Mxy2}) may be used to examine the
implications of specific textures of the neutrino mass matrix.
In the following, we discuss several specific textures of the neutrino
mass matrix which have been considered in the literature.  Their $CP$
effects are particularly noted.

\subsection{$M_{e\mu}=0$}

Specific examples of this class of models are given in Refs.~\cite{bpww},
\cite{gmnr98}, and \cite{roy}, in which $\nu_x = \nu_s$, $\nu_y =
\nu_\tau$, and the mass matrices are taken to be real. In these models
the mass matrices were chosen to minimize
the number of parameters needed to provide the appropriate
phenomenology, and a nonzero $M_{e\mu}$ is not required. In
Ref.~\cite{bpww} the case $\nu_x = \nu_\tau$ and $\nu_y = \nu_s$ was
also considered.

Using Eq.~(\ref{Mem2}), $M_{e\mu}=0$ implies
$s_{12}c_{23} \simeq - s_{13}s_{23}$, which in turn leads to
\bea
A^{e\mu}_{LSND} &\simeq& 16 |s_{13}|^2 |s_{23}|^2 \sin^2\delta_{23} \,,
\label{Aemsbl3}
\\
A^{ey}_{LSND} &\simeq& 4 |s_{13}|^2 (1 - \sin^22\theta_{23}
\sin^2\delta_{23}) / c_{23}^2 \,,
\label{Aeysbl3}
\\
A^{e\mu}_{atm} = - A^{ey}_{atm} &\simeq&
4 |s_{13}|^2 |s_{23}|^2 \cos2\delta_{23} \,,
\label{Aematm3}
\\
B^{e\mu}_{atm} = - B^{ey}_{atm} &\simeq&
-2 |s_{13}|^2 |s_{23}|^2 \sin2\delta_{23} \,,
\label{Bematm3}
\eea
and $\phi_0 = -2\delta_{23}$. Observable oscillations at LSND requires
$\theta_{23} \ne 0, \pi$ and $\delta_{23} \ne 0, \pi$.
Also, $\nu_e \rightarrow \nu_y$ oscillations in short-baseline
experiments, $\nu_e \rightarrow \nu_\mu$ and $\nu_e \rightarrow \nu_y$
oscillations in long-baseline experiments, and $CP$ violation in
long-baseline experiments are possible, although not
required, in this scenario. Finally, we have
\bea
|M_{\mu\mu}| \simeq |M_{yy}| &\simeq&
m \sqrt{1 - \sin^22\theta_{23} \sin^2\delta_{23}} \,,
\label{Mmmmag}
\\
|M_{\mu y}| &\simeq& m \sin2\theta_{23}\sin\delta_{23} \,.
\label{Mmymag}
\eea

As one example, if we take $\delta_{23} \rightarrow {\pi\over2}$ and
$\theta_{23} \rightarrow {\pi\over4}$, we obtain the model of
Ref.~\cite{bpww}, in which there is maximal $\nu_\mu$-$\nu_y$
mixing and $|M_{\mu\mu}| \simeq |M_{yy}| \ll |M_{\mu y}|$. Furthermore
$A^{ey}_{LSND} = B^{e\mu}_{atm} = - B^{ey}_{atm} = 0$, so $\nu_e$
oscillates only to $\nu_\mu$ in short-baseline experiments and there is
no visible $CP$ violation in long-baseline experiments.

If we allow $\theta_{23} \ne {\frac{\pi}{4}}$, we have a model
equivalent to that of Ref.~\cite{gmnr98}; in this case $A^{ey}_{LSND}
\ne 0$ and $B^{e\mu}_{atm} = - B^{ey}_{atm} = 0$, so there can be $\nu_e
\rightarrow \nu_y$ oscillations in short-baseline experiments but still
no visible $CP$ violation in long-baseline experiments.

In order to have $CP$ violation in the present case, we must have
$\delta_{23} \ne {\pi\over2}$. Then there must be short-baseline
$\nu_e \rightarrow \nu_y$ oscillations, although the existence of
long-baseline $\nu_e \leftrightarrow \nu_\mu$ oscillations depends on
the value of $\theta_{23}$.

Finally, an interesting case to consider is maximal $CP$ violation
(maximal in the sense that it gives the largest $CP$-violation
parameter for a given $|s_{13}|$ and $|s_{23}|$), which corresponds to
$\delta_{23} = {\pi\over4}$. If there is also maximal $\nu_\mu$-$\nu_y$
mixing ($\theta_{23}={\pi\over4}$), then the mass matrix in the
$\nu_\mu$-$\nu_y$ sector is approximately (after appropriate changes
of neutrino phase to make the diagonal elements real to leading order in
$\eps$)
\beq
\left( \begin{array}{cc}
M_{\mu\mu} & M_{\mu y} \\
M_{y\mu} & M_{yy} \\
\end{array} \right)
\simeq
{m\over\sqrt2} \left( \begin{array}{cc}
1 & i \\
i & 1 \\
\end{array} \right)
+ {\delta m^2_{atm}\over4m} e^{i\pi/4} \left( \begin{array}{cc}
1 & 1 \\
1 & 1 \\
\end{array} \right) \,.
\label{Mmaxmix}
\eeq
The measurables in short- and long-baseline experiments are then
\bea
A^{e\mu}_{LSND} = A^{ey}_{LSND} &\simeq& 4 |s_{13}|^2 \,,
\label{Aemsbl4}
\\
A^{e\mu}_{atm} = - A^{ey}_{atm} &\simeq& 0 \,,
\label{Aematm4}
\\
B^{e\mu}_{atm} = - B^{ey}_{atm} &\simeq& - |s_{13}^2| \,,
\label{Bematm4}
\eea
i.e., $\nu_e$ oscillates equally into $\nu_\mu$ and $\nu_y$ in
short-baseline experiments and there are no additional contributions
to the $CP$-conserving part of these oscillations in long-baseline
experiments. The vacuum $CP$ and $T$ asymmetries are especially simple
in this case,
\beq
{\cal A}^{CP}_{e\mu} = {\cal A}^{T}_{e\mu}  = - {\cal A}^{T}_{ey}
\simeq - {1\over2} \sin2\Delta_{atm} \,,
\label{ACP2}
\eeq
as the dependence on $|s_{13}|^2$ cancels in the ratio. The particular
models discussed above are summarized in Table~\ref{models}.


\subsection{$M_{ey}=0$}

An example of this class of models with $\nu_x = \nu_s$ and $\nu_y =
\nu_\tau$ is given in Ref.~\cite{mohap}, where a nonzero $M_{ey}$ was
not needed to provide the appropriate phenomenology. From
Eq.~(\ref{Mey2}), $M_{ey}=0$ implies
$s_{12}s_{23}^* \simeq s_{13}c_{23}$, which leads to
\bea
A^{e\mu}_{LSND} &\simeq& 4 |s_{13}|^2 (1 - \sin^22\theta_{23}
\sin^2\delta_{23}) / |s_{23}|^2 \,,
\label{Aemsbl5}
\\
A^{ey}_{LSND} &\simeq& 16 |s_{13}|^2 c_{23}^2 \sin^2\delta_{23} \,,
\label{Aeysbl5}
\\
A^{e\mu}_{atm} = - A^{ey}_{atm} &\simeq&
- 4 |s_{13}|^2 c_{23}^2 \cos2\delta_{23} \,,
\label{Aematm5}
\\
B^{e\mu}_{atm} = - B^{ey}_{atm} &\simeq&
2 |s_{13}|^2 c_{23}^2 \sin2\delta_{23} \,,
\label{Bematm5}
\eea
and $\phi_0 = -2\delta_{23}$. The existence of oscillations in LSND
implies that $\delta_{23} \ne {\pi\over2}$ or $\theta_{23} \ne
{\pi\over4}$. As with the $M_{e\mu}=0$ case, it is possible to have
$\nu_e \rightarrow \nu_y$ oscillations in short-baseline experiments,
$\nu_e \rightarrow \nu_\mu$ and $\nu_e \rightarrow \nu_y$ oscillations
in long-baseline experiments, and $CP$ violation in long-baseline
experiments. The approximate magnitudes of the mass matrix elements
$M_{\mu\mu}$, $M_{yy}$, and $M_{\mu y}$ are the same as given in
Eqs.~(\ref{Mmmmag}) and (\ref{Mmymag}).

The limit $\delta_{23} \rightarrow 0$ and $\theta_{23} \simeq
{\pi\over4}$ reproduces the model in Ref.~\cite{mohap}, which has no
$\nu_e \rightarrow \nu_y$ in short-baseline experiments and no visible
$CP$ violation in long-baseline experiments. CP violation can occur if
$\delta_{23} \ne 0, {\pi\over2}, \pi$, in which case there are $\nu_e
\rightarrow \nu_y$ oscillations in short-baseline experiments and there
may be $\nu_e \rightarrow \nu_\mu$  and $\nu_e \rightarrow \nu_y$
oscillations in long-baseline experiments, depending on the value of
$\theta_{23}$. The $M_{ey}=0$ model with maximal $CP$ violation and
maximal $\nu_\mu$-$\nu_y$ mixing ($\delta_{23} = \theta_{23} =
{\pi\over4}$) has the same features as the $M_{e\mu}=0$ maximal $CP$
violation case in Sec.~IV.A, except that the vacuum $CP$ and $T$
asymmetries in Eq.~(\ref{ACP2}) have the opposite sign. The particular
$M_{ey}=0$ cases discussed here are also summarized in
Table~\ref{models}.

\subsection{$M_{e\mu} \ne 0$ and $M_{ey} \ne 0$}

In this more general case, barring fortuitous cancellations one would
expect from Eqs.~(\ref{Aemsbl2})--(\ref{Bmxatm2}) that there are
$\nu_e \rightarrow \nu_\mu$ and $\nu_e \rightarrow \nu_y$ oscillations
in short- and long-baseline experiments, and $CP$ violation in $\nu_e
\rightarrow \nu_\mu$ and $\nu_e \rightarrow \nu_y$ oscillations in
long-baseline experiments.

For this texture, $|M_{\mu\mu}|, |M_{yy}| \ll |M_{\mu y}|$ does
not necessarily exclude visible $CP$ violation, unlike the cases
$M_{e\mu}=0$ or $M_{ey}=0$. As an example, the mass matrix
\beq
M = m \left( \begin{array}{cccc}
\eps_1 & \eps_2 e^{i\phi_2} & 0 & 0 \\
\eps_2 e^{i\phi_2} & 0 & \eps_5 & \eps_3 e^{i\phi_3} \\
0 & \eps_5 & \eps_4 & e^{i\phi_1} \\
0 & \eps_3 e^{i\phi_3} & e^{i\phi_1} & \eps_6 \\
\end{array} \right) \,,
\label{M}
\eeq
which is an extension of the model introduced in Ref.~\cite{bpww},
leads to $CP$ violation of order $\eps^2$ in $\nu_e \rightarrow \nu_\mu$
and $\nu_e \rightarrow \nu_y$ oscillations in long-baseline experiments.
This mass matrix differs from the one in Ref.~\cite{bpww} in that the
$M_{e\mu}$ and $M_{\mu e}$ elements, denoted as $\eps_5$, are not zero,
the $M_{\mu\mu}$ and $M_{yy}$ elements, denoted as $\eps_4$ and
$\eps_6$, respectively, are not necessarily equal, and the
$CP$-violating phases are not set to zero. The diagonal elements of the
mass matrix can be taken to be real.  Because
$M_{ee} = M_{x\mu} = M_{xy} = 0$ there are only three independent
phases. The mass eigenvalues and approximate mixing matrix for the mass
matrix in Eq.~(\ref{M}) are given in Appendix~C.

The largest off-diagonal short-baseline oscillation amplitudes in this
case are
\bea
A^{e\mu}_{LSND} &\simeq& 4 \eps_3^2 \,,
\label{Aemsbl6}
\\
A^{ey}_{LSND} &\simeq& 4 \eps_5^2 \,.
\label{Aeysbl6}
\eea
Short-baseline $\nu_\mu$-$\nu_y$ oscillations are of order $\eps^4$.
The largest long-baseline oscillation probabilities are
\bea
A^{\mu y}_{atm} &\simeq& 1 \,,
\label{Amyatm6}
\\
A^{e\mu}_{atm} = - A^{ey}_{atm} &\simeq& \eps_5^2 - \eps_3^2 \,.
\label{Aematm6}
\eea
Short- and long-baseline oscillation amplitudes involving $\nu_x$ are of
order $\eps^4$ or smaller. The observable $CP$-violating amplitude is
\beq
B^{e\mu}_{atm} \simeq \eps_3 \eps_5 \sin(\phi_3 + \delta_{23}) \,,
\label{Bematm6}
\eeq
where (see Appendix~C)
\beq
\tan \delta_{23} = - {(\eps_4 - \eps_6)\sin\phi_1 + 2\eps_3\eps_5\sin\phi_3
\over (\eps_4 + \eps_6)\cos\phi_1 + 2\eps_3\eps_5\cos\phi_3 } \,.
\label{delta23}
\eeq
Finally,
\beq
A^{ee}_{sun} = \sin^2 \theta_{01} \,,
\label{Aeesun6}
\eeq
where
\beq
\tan \theta_{01} = {\eps_2 \over \eps_1 + 2 \eps_3\eps_5}
\sqrt{1 + {4\eps_1\eps_3\eps_5(1-\cos(\phi_1 +\phi_3 - 2\phi_2))
\over (\eps_1 - 2\eps_3\eps_5)^2} } \,.
\eeq
The phenomenology of these models is summarized in Table~\ref{models}.

\subsection{$M_{ee} = M_{e\mu} = M_{\mu\mu} = 0$}

In the special class of models with two sterile and two active
neutrinos, i.e., both $\nu_x$ and $\nu_y$ are sterile, there need not be
Majorana mass terms for the two active neutrinos in order to obtain the
proper phenomenology. As described in more detail in Appendix~B, $M_{ee}
= M_{e\mu} = M_{\mu\mu} = 0$ requires only a minimal extension in the
Higgs sector of the Standard Model, i.e., only $SU(2)$ singlet Higgs
bosons need to be added.  Examples of models with both $\nu_x$ and
$\nu_y$ sterile are given in Ref.~\cite{krolikowski}. We will now show
that $CP$ violation effects in long-baseline experiments are
suppressed in this class of models.

Models with $M_{e\mu}=0$ have already been discussed in Sec.~IV.A; here
we add the additional constraint $M_{\mu\mu}=0$. It was previously
determined that $|M_{\mu\mu}|, |M_{yy}| \ll |M_{\mu y}|$ implied
$\delta_{23} \simeq {\pi\over2}$ and $\theta_{23} \simeq {\pi\over4}$,
so Eqs.~(\ref{Aemsbl3})--(\ref{Bematm3}) reduce to
\bea
A^{e\mu}_{LSND} &\simeq& 8 |s_{13}|^2 \,,
\label{Aemsbl7}
\\
A^{ey}_{LSND} &\simeq& 0 \,,
\label{Aeysbl7}
\\
A^{e\mu}_{atm} = - A^{ey}_{atm} &\simeq& - 2 |s_{13}|^2 \,,
\label{Aematm7}
\\
B^{e\mu}_{atm} = - B^{ey}_{atm} &\simeq& 0 \,.
\label{Bematm7}
\eea
Hence, there is no visible $CP$ violation in long-baseline experiments
in this case (strictly speaking $CP$ violation is strongly suppressed,
to order $\eps^4$). Therefore in models with $M_{ee} = M_{e\mu} =
M_{\mu\mu} = 0$, the only phenomenological deviations from the Standard
Model are $CP$-conserving neutrino oscillations (see
Table~\ref{models}) and the presence of more than one neutral Higgs
scalar.

\subsection{Other mass hierarchies}

There are other mass hierarchies possible which give the same
oscillation phenomena as those discussed in Secs.~IV.A--IV.D. One is
$m_2 < m_3 \ll m_0, m_1 \simeq
\sqrt{\delta m^2_{LSND}} \equiv m$, in which the solar oscillation
occurs between the two upper mass eigenstates and the atmospheric
oscillation between the two lower mass eigenstates. Assuming as before
that $\nu_e$ mixes primarily with $\nu_x$ and $\nu_\mu$ with $\nu_y$
(i.e., $s_{02}, s_{03}, s_{12}, s_{13} \sim \eps$), then
\beq
M_{e\mu} \simeq m (U_{e0}^* U_{\mu0}^* + U_{e1}^* U_{\mu1}^*) \,;
\label{Mem3}
\eeq
$M_{ee}$, $M_{ex}$, $M_{xx}$, $M_{ey}$, $M_{x\mu}$, and $M_{xy}$ are
given by similar expressions with appropriate changes of subscripts.
However, for $M_{\mu\mu}$, $M_{\mu y}$, and $M_{yy}$ none of the four
terms in Eq.~(\ref{Mab}) are dominant. Since long-baseline $\nu_\alpha
\rightarrow \nu_\beta$ oscillations involve $\delta m^2_{32}$, and hence
the mixing matrix elements $U_{\alpha2}$, $U_{\alpha3}$, $U_{\beta2}$,
and $U_{\beta3}$ (see Eqs.~(\ref{Ammatm})--(\ref{Bmyatm})), then a mass
texture condition such as $M_{e\mu}=0$, when applied to
Eq.~(\ref{Mem3}), does not tell us anything specific about
long-baseline oscillations. It could, however, affect short-baseline
oscillation amplitudes, which depend on the $U_{\alpha0}$ and
$U_{\alpha1}$ (see Eqs.~(\ref{Aemsbl})--(\ref{Aeesbl})). We do not
pursue this possibility further here.

Another possible hierarchy is to have $m_0 \simeq m_1 < m_2 \simeq m_3$
where none of the masses are much smaller than the others; in this case,
all masses would contribute to hot dark matter (an alternate possibility,
$m_2 \simeq m_3 < m_0 \simeq m_1$ with none small, gives similar
results). A model of this type has been discussed in Ref.~\cite{roy}.
In this case, again assuming $s_{02}, s_{03}, s_{12}, s_{13} \sim \eps$,
we find
\beq
M_{ex} \simeq m_0 (U_{e0}^* U_{x0}^* + U_{e1}^* U_{x1}^*) \,,
\label{Mex4}
\eeq
with similar expressions for $M_{ee}$ and $M_{xx}$, and
\beq
M_{\mu y} \simeq m_3 (U_{\mu2}^* U_{y2}^* + U_{\mu3}^* U_{y3}^*) \,,
\label{Mmy4}
\eeq
with similar expressions for $M_{\mu\mu}$ and $M_{yy}$. However, for
$M_{e\mu}$, $M_{ey}$, $M_{x\mu}$, and $M_{xy}$, none of the four terms
in Eq.~(\ref{Mab}) are dominant, the expressions for the mass matrix
elements are more complicated, and the implications of particular
textures for long-baseline oscillations are not as easily
determined. We also do not pursue this case further here.

\section{Other probes of neutrino mass}

The presence of mass terms for neutrinos and, in particular, Majorana
mass terms, opens up a variety of possibilities for phenomena that are
not possible in the Standard Model.  The neutrino mass, besides giving
rise to mixing of neutrinos and the associated $CP$ effect discussed in
this paper, can lead to lepton flavor-changing charged currents
analogous to those of the quark sector. Majorana mass can also give rise
to lepton number violation processes.  With these possibilities, widely
searched-for phenomena such as $\mu \rightarrow e + \gamma$,
$\mu \rightarrow e + \bar{e} + e$, $\mu$-$e$ conversion, and electric
dipole moments for charge leptons, can occur.  Unfortunately, all
these processes \cite{textbooks} are proportional to $(m_\nu / M_W)^4$
or $((m_\nu/M_W)ln(m^2_\nu/M^2_W))^4$.  Given that $m_\nu$ is of the
order of 5~eV or less~\cite{bww} these are no larger than $10^{-40}$ and
$10^{-33}$.  The current upper bounds \cite{pdg} are about 30 orders of
magnitude larger than those theoretical predictions from the neutrino
masses.  Therefore they are unobservable. Since this conclusion
depends only on the smallness of the neutrino masses, it is valid in
general.

The Majorana mass term breaks lepton number conservation and can
lead to neutrinoless double beta decay. The rate is governed
by the magnitude of the effective $\nu_e$ mass 
\beq
\langle m_{\nu_e} \rangle = |\sum_j U_{ej}^2 m_j| = |M_{ee}| \,,
\eeq
i.e., the magnitude of the $M_{ee}$ element in the Majorana
neutrino mass matrix in Eq.~(\ref{Mflavor}). The current limit on
$|M_{ee}|$ from neutrinoless double beta decays is about
0.5~eV~\cite{KK}. Since
\beq
M_{ee} =
s_{01}^{*2} m_0 + c_{01}^2 m_1 + s_{12}^{*2} m_2 + s_{13}^{*2} m_3 \,,
\label{Mee2}
\eeq
there will be no visible neutrinoless double beta decay in models with
$m_0, m_1 \ll m_2 < m_3 \simeq \sqrt{\delta m^2_{LSND}} \approx 1$~eV
and $|s_{12}| \approx |s_{13}| \approx \eps$. In models where $m_2 < m_3
\ll m_0, m_1 \simeq \sqrt{\delta m^2_{LSND}}$, or if no neutrino masses
are $\ll 1$~eV, neutrinoless double beta decay may provide a strong
constraint.

Neutrino masses may also affect cosmology if $\sum_\nu m_\nu >
0.5$~eV~\cite{cosmology}. This level of neutrino mass can easily be
accommodated by a four-neutrino model with two pairs of nearly
degenerate mass eigenstates separated by approximately 1~eV.

Because of the smallness of the neutrino masses, there are no other
observable effects besides neutrino oscillations and possibly
neutrinoless double beta decay and dark matter. However, the rare decays
may still be observable if new physics occurs also in other sectors,
such as anomalous gauge boson couplings or anomalous fermion-gauge boson
interactions.  Therefore, it is important to continue to search for
them.

\section{Summary}

In this paper we have presented a general parametrization of the
four-neutrino mixing matrix and discussed the oscillation phenomenology
for the case of two nearly degenerate pairs of mass eigenstates
separated from each other by approximately 1~eV, which is the mass
spectrum indicated by current solar, atmospheric, reactor, and
accelerator neutrino experiments. We analyzed in detail the case where
$\nu_e$ mixes primarily with $\nu_x$ and $\nu_\mu$ with $\nu_y$, where
one of $\nu_x$ and $\nu_y$ is $\nu_\tau$ and the other is sterile, or
both are sterile. We found in these cases that the neutrino mixing
matrix can be written in $2\times2$ block form with small off-diagonal
blocks. By construction the mixing matrices have $\nu_e \rightarrow
\nu_\mu$ oscillations in LSND and $\nu_\mu \rightarrow \nu_y$
oscillations in atmospheric experiments. We found that the following
oscillations are also possible: $\nu_e \rightarrow \nu_y$ and $\nu_\mu
\rightarrow \nu_x$ in short-baseline experiments, and $\nu_e
\rightarrow \nu_\mu$, $\nu_e \rightarrow \nu_y$, and $\nu_\mu
\rightarrow \nu_x$ (including $CP$-violation effects) in long-baseline
experiments. We also found that solar, atmospheric, short- and
long-baseline oscillation measurements can, in some cases, determine
all but one of the four-neutrino mixing parameters. Finally, we
examined the implications of some several specific mass textures, and
found the conditions under which $CP$-violation effects are visible.

As pointed out in the Introduction, additional evidence is needed in
order for the four-neutrino scenario to be on firm ground. Given that
there must be separate mass-squared difference scales for the solar and
atmospheric oscillations (as currently indicated by the data), there are
in fact two ways to verify the existence of four light neutrinos: (i)
confirmation of the LSND results, which could occur in the future
mini-BOONE collaboration \cite{miniBOONE,Conrad,nui}, or (ii) detection
of vacuum (i.e., not matter-induced) $CP$ or $T$ violation in
long-baseline experiments, which should be greatly suppressed in a
three-neutrino scenario.

Once the existence of four neutrinos is established, the next task is to
determine the neutrino mixing matrix parameters. We emphasize the
significant potential for detecting new oscillation channels and $CP$
violation in future high statistics short- and long-baseline
oscillation experiments. Many experiments have been proposed and some
will be online in the next few years~\cite{nui,geer}.  In these
experiments neutrino beams are produced at high energy accelerators and
oscillations can be detected at distant underground detectors. They
include the KEK-Kamiokande K2K Collaboration \cite{K2K}, the
Fermilab-Soudan MINOS \cite{MINOS} and Emulsion Sandwich \cite{emulsion}
collaborations, and CERN-Gran Sasso ICARUS, Super-ICARUS, AQUA-RICH,
NICE, NOE and OPERA collaborations~\cite{GranSasso}. Experiments done at
muon storage rings~\cite{geer} may be especially important since they
will have the ability to measure both $\nu_e \rightarrow \nu_\mu$ and/or
$\nu_\tau$, and $\nu_\mu \rightarrow \nu_e$ and/or $\nu_\tau$, as well
as the corresponding oscillation channels for antineutrinos.
Furthermore, there may also be hitherto undiscovered oscillation effects
in short-baseline oscillation experiments such as COSMOS~\cite{COSMOS}
and TOSCA~\cite{TOSCA}, which will search for $\nu_\mu \rightarrow
\nu_\tau$ oscillations. To completely determine all accessible
parameters in the four-neutrino mixing matrix requires searches at both
short and long baselines.  The amplitudes of various oscillation
channels, including possible $CP$ violation effects, will help further
determine the texture of the four-neutrino mass matrix and offer a
better understanding of neutrino physics as well as $CP$ violation.

\bigskip
\section*{Acknowledgements}

YD would like to thank the hospitality of Iowa State University where this
work began.  This work was supported in part by the U.S. Department of
Energy, Division of High Energy Physics, under Grants
No.~DE-FG02-94ER40817 and No.~DE-FG02-95ER40896, and in part by the
University of Wisconsin Research Committee with funds granted by the
Wisconsin Alumni Research Foundation.
YD's work is partially support by the National Natural Science
Foundation of China.
BLY acknowledges the support by a NATO collaborative grant.

\appendix
\section{General properties of $n\times n$ Majorana mass and
mixing matrices}

Consider the general case of $n$ neutrinos, in which $n_R$ are
right-handed (sterile) and $n_L$ are left-handed (active),
$n_R+n_L=n$.  We can represent the $n_R$ right-handed neutrinos by
their left-handed conjugates and denote the collection of all the $n$
independent left-handed neutrinos by a column vector $\psi_L$.  Then
either Dirac or Majorana neutrino mass terms can be written as
\beq
\bar\psi_R M \psi_L + h.c. \,,
\label{Majorana}
\eeq
with $\bar\psi_R$ related to $\psi_L$ by
$\bar\psi_R = \psi^T_{L}C$, where $C = i\gamma^2\gamma^0$
is the charge conjugation matrix.
The most general $n\times n$ mass matrix $M$ is symmetric and
complex, $M^\dagger = M^*$, characterized by $\half n(n+1)$ magnitudes and
the same number of phases.  Since a given component field in $\psi_L$ and
$\bar\psi_R$ acquires the same phase factor under a change of phase, $n$
of the phases in $M$ may be absorbed into the definitions of the fields,
leaving $\half n(n+1)$ magnitudes and $\half n(n-1)$ phases; we will often
choose the convention that the diagonal elements of $M$ are real and the
off-diagonal elements complex.

As shown in Eq.~(\ref{diag}), the mass matrix may be diagonalized by a
unitary matrix $U$. The effects of $U$ may be divided into two classes
of phenomenology which give rise to violation of individual lepton
number: neutrino oscillations and lepton charged currents. Since $Z^0$
interacts only with the left-handed neutrinos, neutrino counting in $Z
\rightarrow \nu \bar\nu^\prime$ at LEP is unchanged if all of the
neutrinos are light, i.e., $(N_\nu)_{LEP} = n_L$, with $n_L = 3$ in the
Standard Model. This can be demonstrated straightforwardly as
follows. The neutral current Lagrangian can be written in terms of
flavor eigenstates as $g\bar{\psi}_L \gamma^\mu K \psi_L Z^0_\mu$, where
$K$ is an $n\times n$ diagonal matrix with the first $n_R$ elements
being zero and the other $n_L$ elements unity. If all neutrinos are very
light, which is the case we are considering here, the number of
neutrinos measured at LEP is simply Tr($K{K^\dag}) = n_L$, assuming
the couplings of the active neutrinos are universal.  In terms of the
mass eigenstates, the neutrino counting is unchanged:
Tr($UK{(UK)^\dag}) = {\rm Tr}(K{K^\dag}) = n_L$.

In general an $n\times n$ unitary matrix such as $U$ can be described by
${1\over2}n(n-1)$ rotation angles and ${1\over2}n(n+1)$ phases. In the
charged lepton current, we are free to make phase transformations of the
charged lepton fields, which removes $n$ of the phases. Then the number
of surviving measurable phases is ${1\over2}n(n+1) - n =
{1\over2}n(n-1)$.  This argument is not affected by the fact that
the number of left-handed charged lepton fields, $n_c$, in the
charged-current is less than $n$, as each of the first $n-n_c$ rows of
$U$ can be multiplied by a phase without altering the charged currents.
Therefore, in general in this Majorana setting we can parametrize $U$ by
${1\over2}n(n-1)$ angles and ${1\over2}n(n-1)$ phases.

In neutrino oscillations, however, only ${1\over2}(n-1)(n-2)$
independent phases can in principle be measured, as we will now
demonstrate. Note that the $W^{jk}_{\alpha\beta}$ in Eq.~(\ref{W}) are
invariant when $U$ is transformed from either the left or right side by
a diagonal matrix which contains only phases, i.e.,
\beq
U_{\alpha j} \rightarrow e^{i\phi_\alpha} U_{\alpha j} e^{i\phi_j} \,.
\label{phase}
\eeq
Then, {\it as far as neutrino oscillations are concerned}, we can
eliminate $2n-1$ of the phases in $U$, so that there are effectively
only
\beq
{1\over2}n(n+1) - (2n-1) = {1\over2}(n-1)(n-2) \,,
\label{phasenumber}
\eeq
independent phases that can be measured by neutrino oscillation
experiments. Interestingly this is the same number of independent
phases that may be determined in the CKM matrix for $n$ generations of
quarks. However, note that in the most general $U$ there are
${1\over2}n(n-1)$ phases, so that there are ${1\over2}n(n-1) -
{1\over2}(n-1)(n-2) = (n-1)$ phases that cannot be determined from
neutrino oscillations.

The phase counting of Eq.~(\ref{phasenumber}) can also be confirmed by
enumerating the number of independent $CP$-violating variables, e.g.,
the $\Delta P_{\alpha\beta}$ defined in Eq.~(\ref{deltap}), that can be
measured. There are ${1\over 2}n(n-1)$ such differences, but from
Eq.~(\ref{Jasy}) $\Delta P_{\alpha\beta} = - \Delta P_{\beta\alpha}$ and
from Eq.~(\ref{Jsum}) $\sum_\beta \Delta P_{\alpha\beta} = 0$, so it
follows that $n-1$ of the differences are not independent. Therefore
there are only ${1\over2}(n-1)(n-2)$ independent $\Delta
P_{\alpha\beta}$, and only ${1\over2}(n-1)(n-2)$ $CP$-violation
parameters can be measured.

\section{Higgs boson origins of neutrino masses}

The presence of masses for neutrinos is a definite signal of physics
beyond the Standard Model.  Particularly, with the three types of
neutrino oscillations which indicate three $\delta m^2$ scales and
require at least four neutrino mass values, a non-trivial extension of
the Standard Model is necessary.  In searching for hints of the
extension, it is interesting to consider what simplest extensions of the
Standard Model are possible and how natural (or unnatural) they are in
their couplings schemes.  In this appendix we discuss briefly the
possible origins of the two types of mass matrices considered, i.e.,
models with one or two sterile neutrinos. Then both Dirac
(active-sterile) and Majorana (active-active or sterile-sterile)
neutrino mass terms are present.  These masses can be obtained by
suitable extensions of the Standard Model. We will only enlarge the
lepton Yukawa sector and the Higgs sector to the extent required by the
mass matrices of Sec.~IV.  Our goal is to assure that such mass matrices
are possible by straightforward modifications of these two sectors, and
to determine what new particles need to be added to the Standard Model
spectrum. We do not attempt to construct the best case scenario, which
can be done when more information on the neutrinos mass are available.
For a more extensive discussion of possible origins of neutrino masses
terms, see Ref.~\cite{Gelmini}.

In the following we denote the right-handed sterile by $\nu_{sjR}$
and the corresponding left-handed conjugate by
$\hat{\nu}_{sjL}=\bar{\nu}^T_{sjR} C$.  We also denote the
left-handed lepton SU(2) doublet by $l_{kL}$ with the corresponding
right-handed conjugate fields
$\hat{l}_{kR}=-i\sigma_2C\bar{l}^T_{kR}$, $j$ and $k$ are generation
labels.

\subsection{Models with two left-handed and two right-handed neutrinos}

We first consider the class of models in which $\nu_x = \nu_{s1}$ and
$\nu_y = \nu_{s2}$, where $\nu_{s1}$ and $\nu_{s2}$ are the
right-handed sterile neutrinos that are associated with $\nu_e$ and
$\nu_\mu$, respectively, which have been considered in
Ref.~\cite{krolikowski}. We discuss two cases: (i) models where only the
sterile neutrinos have nonzero Majorana mass terms, and (ii) models
where both right-handed (sterile) and left-handed ($\nu_e$ and
$\nu_\mu$) fields have Majorana mass terms. In the first case $M_{ee} =
M_{e\mu} = M_{\mu\mu} = 0$, which is discussed in Sec.~IV.D; in the
second case, $M_{e\mu} \ne 0$, which can be realized in the models
discussed in Secs.~IV.B and IV.C.

The simplest extension of the Standard Model is case (i) above, which
can be obtained by adding a singlet real
scalar field $\phi$ to the Standard Model Higgs doublet $\Phi$.  The
Majorana masses for the two sterile are due to their coupling to $\phi$
and are proportional to the vacuum expectation value (vev) of
$\phi$. The $\Phi$ provides the Dirac masses from Yukawa couplings
involving both sterile and left-handed neutrinos.  We denote the
absolute value of the vev's of the $\Phi$ and $\phi$ fields as $v$ and
$V$ respectively; $v$ is the same as the Standard Model vev.  The Yukawa
couplings can be written as
\beq
 {\cal{L}}_Y = \sum_{j,j'}^{2} G_{jj'}\phi \bar{\nu}_{sjR}\hat{\nu}_{sj'L}
             + \sum_{j,k} g_{jk}(\bar{\nu}_{sjR}{\tilde{\Phi}^\dag}l_{kL}
                                + \bar{\hat{l}}_{kR}\Phi\hat{\nu}_{sR})
            + h.c. \,,
\eeq
where $G_{jj'}$ and $g_{jk}$ are complex couplings, and
$\tilde{\Phi} = i\sigma\Phi^*$.  Since
$\bar{\nu}_{sjR}\hat{\nu}_{sj'L} = \hat{\nu}_{sj'L}\bar{\nu}_{sjR}$ we
have $G_{jj'} = G_{j'j}$.  We also exhibit the symmetry of the Dirac
coupling coefficients, $g_{jk}=g_{kj}$, because of the identity
$\bar{\nu}_{sjR}{\tilde{\Phi}^\dag}l_{kL}
  = \bar{\hat{l}}_{kR}\Phi\hat{\nu}_{sjR}$

For the Higgs potential, we take the simplified case that it has a $Z_2$
symmetry in $\phi$, i.e., invariant under $\phi \rightarrow -\phi$.  Then the
Higgs potential contains only real coefficients:
\beq
 {\cal{L}_H} = -\mu^2_1 |\Phi|^2 - \mu^2_2 \phi^2 + \lambda |\Phi|^4
               + \lambda_2\phi^4 + \lambda_3 \phi^2 |\Phi|^2 \,.
\eeq
After spontaneous symmetry breaking there are two massive neutral Higgs
bosons. Their masses are set by $v$ and $V$.  The value of $v$ is the same
as in the Standard Model.  The coefficients of the
Yukawa couplings must be very small in order to give neutrino masses
of the order of eV.  In the case $V \gg v$, one of the Higgs boson is composed
mostly of the neutral field of $\Phi$ with a mass proportional to $v$ and
the other composed of mostly the $\phi$ field with a mass proportional to
$V$.  There are no other changes to the Standard Model phenomenology.

One can also have a more complicated scenario of case (ii) in which the
Majorana masses of the two left-handed neutrinos are non-vanishing.
These types of models can be constructed by the approach discussed below
in Appendix~B.2.

\subsection{Models with three left-handed and one right-handed
neutrinos}

Here we consider the cases (i) $\nu_x = \nu_s$ and $\nu_y = \nu_\tau$, or
(ii) $\nu_x = \nu_\tau$ and $\nu_y = \nu_s$, where $\nu_s$ is a sterile
neutrino. In the first case, mass terms are needed in the
$\nu_\mu$-$\nu_\tau$ sector to provide the large mixing of atmospheric
neutrinos, while in the second case Majorana mass terms are needed to
provide mixing of solar neutrinos. There are no constraints on which
terms in the mass matrix may be nonzero, but in each case Majorana
masses of the left-handed (active) neutrinos must exist. The
phenomenology of some of these models is discussed in Secs.~IV.A--IV.C.

Majorana mass terms between left-handed neutrinos can arise from the
introduction of a Higgs triplet which has lepton number
$-2$~\cite{GelminiRoncadelli}. The Majorana mass of the sterile neutrino
again comes from a Higgs singlet as discussed above.
We denote the triplet by $\Delta$
\beq
\Delta = {1\over\surd{2}} \vec{\tau} \cdot \vec{\delta}
       = \left( \begin{array}{cc}
                 \delta^+/\surd{2}  &  \delta^{++} \\
                 \delta^0          &  -\delta^+/\surd{2}
                 \end{array} \right) \,,
\eeq
and the value of the vev of
$\delta^0$ by $\Lambda$.  Since $\Lambda$ contributes to the masses of
the W and Z$^0$ bosons differently, it has to be small in comparison
with the vacuum expectation of the $\Phi$, say $\Lambda/v < 10^{-2}$,
so that the bulk of the electroweak gauge boson masses come from the
doublet.  Then, this will not upset the good agreement achieved by
the Standard Model prediction for the $\rho$ parameter.

The Yukawa couplings can be written as
\beq
 {\cal{L}}_Y = G_s\phi \bar{\nu}_{sR}\hat{\nu}_{sL}
             + \sum_k g_k(\bar{\nu}_{sR}{\tilde{\Phi}^\dag} l_{kL}
                        +\bar{\hat{l}}_{kL}{\Phi} {\nu}_{sR})
             + \sum_{k,k'}h_{kk'}\bar{\hat{l}}_{kR}\Delta l_{k'L}
             + h.c. \,,
\label{Yuk}
\eeq
where $G_s$, $g_k$ and $h_{kk'}$, $k$ and $k'=1,2,3$, are complex
couplings.  The symmetry of the Dirac couplings is explicitly exhibited
because of the identity
$\bar{\nu}_{sR}{\tilde{\Phi}^\dag} l_{kL}
                      = \bar{\hat{l}}_{kL}{\Phi} {\nu}_{sR}$.
$h_{kk'}$ is symmetric, $h_{kk'} = h_{k'k}$, because of the identity
$\bar{\hat{l}}_{kR}\Delta l_{k'L} = \bar{\hat{l}}_{k'R}\Delta l_{kL}$.
For the Higgs potential, we can again take the simplified case that
$\phi$ is a real scalar field and the Higgs potential has the $Z_2$
symmetry in $\phi$:
\bea
 {\cal{L}}_H = &-&\mu^2_1 |\Phi|^2 - \mu^2_2 \phi^2
   -\mu^2_3{Tr(\Delta \Delta^\dag)}+\eta {\Phi^\dag} \Delta {\tilde{\Phi}}
   + \eta^*{\tilde{\Phi}^\dag}\Delta^\dag \Phi
   + \lambda_1|\Phi|^4 + \lambda_2\phi^4
\nonumber\\
   &+&\lambda_3(Tr(\Delta{\Delta^\dag}))^2
   +\lambda_4 {Tr((\Delta\Delta^\dag)^2)}
   +\lambda_5 {Tr(\Delta^2({\Delta^\dag})^2)}
   +\lambda_6 {Tr{\Delta^2}}{Tr{((\Delta^\dag)^2)}}
\nonumber\\
   &+& \xi_1\phi^2|\Phi|^2 + \xi_2 \phi^2{Tr(\Delta{\Delta^\dag})}
   + \xi_3 {\Phi^\dag}\Delta{\Delta^\dag}\Phi
   + \xi_4 {\Phi^\dag}{\Delta^\dag}\Delta \Phi
   + \xi_5 |\Phi|^2 Tr(\Delta{\Delta^\dag})\,.
\label{Lag}
\eea
The following terms in the Higgs potential,
\beq
\eta \Phi^\dagger \Delta \tilde{\Phi}
     + \eta^* \tilde{\Phi}^\dagger \Delta^\dagger \Phi \,,
\label{tri}
\eeq
are needed to break the global lepton number invariance in order to avoid
the appearance of a Goldstone boson called Majoron
\cite{GelminiRoncadelli}, and $\eta$ is the only coupling that
potentially can be complex.

To obtain the neutrino mass matrix we can also make phase tranformations
on the fermion fields $\nu_{sR}$ and $l_{kL}$ to make $G_s$ and the
diagonal terms $h_{kk}$ real. If $CP$ is not broken spontaneously, which
we assume to be the case here, the vacuum expectation values of all
neutral fields can be made real by phase transformations on the Higgs
fields $\Phi$ and $\Delta$.  After spontaneous symmetry breaking this
choice of phases for the Yukawa couplings agrees with the convention of
the neutrino mass matrix discussed in Sec.~II.

With complex couplings, $CP$ violation can generally occur in the Higgs
sector. However, $\eta$ is required to be real by the minimization of
the Higgs potential. Hence explicit $CP$ violation does not occur in
this extended Higgs scenario.  A more complicated Higgs potential can be
chosen to allow complex couplings so that $CP$ violation can be manifest
in the Higgs sector. We will not elaborate on this possibility here.

After spontaneous symmetry breaking, the physical Higgs boson spectrum
contains a doubly charged pair, a singly charged pair, and four neutrals.
The Goldstone bosons are mostly from the Higgs doublet $\Phi$. The masses of
two of the neutral Higgs bosons are proportional to $v$.  The masses of
the remaining two neutral Higgs boson are similar to those of the case
of Appendix~B.1.  Again the Majorana couplings of the sterile neutrino
and the Dirac coupling of the sterile to the left-handed neutrinos are
small.  However, the Majorana couplings among the left-handed leptons do
not have to be small if $\Lambda$ is chosen to be the order of the
neutrino masses, i.e., eV~\cite{Langacker}; this can be done without
leading to any small Higgs boson masses.  Then the coupling of the singly
and doubly charged Higgs boson to the charged leptons are not small.  The
production of these particles in a future high energy linear collider
or muon collider is possible if they are not too heavy.

In this type of model, since the constraints $M_{ee} = M_{e\mu} =
M_{\mu\mu} = 0$ do not apply, the $CP$-violating parameter
Im$(U_{e2}U_{e3}^*U_{\mu2}^*U_{\mu3})$ is no longer constrained to be
approximately zero. The $CP$ violation can be of order $\eps^2$, which
is the same order as the oscillation probabilities themselves, and hence
measurable.  Examples of this type of model include the maximal $CP$
violation models characterized by Eq.~(\ref{Mmaxmix}) and the mass
matrix in Eq.~(\ref{M}).

We note that to produce the required neutrino masses and the
phenomenologically interesting mass spectrum and expectation values of
the Higgs boson fields in both models dicussed in this section, new
hierarchy problems are introduced~\cite{Langacker}.  In our view, such
hierarchy problems do not necessarily argue against the models. However,
it does argue that any model of this sort should be included in a
larger, more natural, scheme. Note that although the hierarchy in the
expectation values of the Higgs boson fields sometimes leads to a fine
tuning of the parameters, the small vacuum expection value of $\Delta$
may be obtained in a natural way~\cite{masarkar}. For example, if $\mu_3
\gg v^2, V^2, \eta^2 \gg \Lambda^2$, then the minimization of the Higgs
potential leads to the relation $\Lambda \simeq - \eta v^2/\mu_3 \ll v$.

There is a growing literature on the generation of neutrino masses. An
intriguing class of models are those that generate mass dynamically by
higher order loop effects~\cite{zee}. We refer the reader to
Ref.~\cite{rad} for recent and extensive analyses of this possibility
for four neutrinos. There are also models that use lepton-number
violating interactions in $R$-parity violating supersymmetry for the
generation of Majorana mass~\cite{RPV}.

\section{An example with $CP$ violation}

In this appendix we derive the masses and mixing matrix for the model
described by the mass matrix in Eq.~(\ref{M}). In general a $4\times4$
Majorana mass matrix can have six independent phases (see Appendix~A),
but since three of the mass matrix elements are zero, there are only
three independent phases in this case. We have chosen to make the
diagonal elements of $M$ real.

To achieve the proper neutrino phenomenology, we assume the following
hierarchy
\beq
\eps_2 \ll \eps_1,\eps_4,\eps_6 \ll \eps_3,\eps_5 \ll 1 \,.
\label{hierarchy}
\eeq
The mass-squared eigenvalues are approximately given by
\beq
m_0^2 \simeq \eps_1^2 m^2 \,, \qquad
m_1^2 \simeq 4\eps_3^2\eps_5^2 m^2 \,, \qquad
m_{2,3}^2 \simeq (1 + \eps_3^2 + \eps_5^2 \mp \eps_0^2) m^2 \,,
\label{eigval}
\eeq
where
\beq
\eps_0^4 = 4\eps_3^2\eps_5^2 + (\eps_4-\eps_6)^2 + 4\eps_4\eps_6 c_1^2
+4\eps_3\eps_5 [ \eps_4\cos(\phi_1-\phi_3) + \eps_6\cos(\phi_1+\phi_3) ] \,,
\label{m0}
\eeq
with $c_j \equiv \cos\phi_j$ and $s_j \equiv \sin\phi_j$. The
eigenvalues are related to the physical mass-squared differences by
\bea
\delta m^2_{LSND} &=& m_2^2 - m_1^2 \simeq m^2 \,,
\label{dm2LSND}
\\
\delta m^2_{atm} &=& m_3^2 - m_2^2 \simeq 2\eps_0^2 m^2 \,,
\label{dm2atm}
\\
\delta m^2_{sun} &=& m_0^2 - m_1^2 \simeq (\eps_1^2 - 4\eps_3^2\eps_5^2) m^2
\,.
\label{dm2sun}
\eea
The size of $\delta m^2_{sun}$ in Eq.~(\ref{dm2sun}) implied by the
hierarchy of Eq.~(\ref{hierarchy}) means that the solar neutrino
oscillations are of the MSW type. Therefore in order to have the proper
MSW enhancement in the sun we must have $m_0^2 > m_1^2$, which implies
$|\eps_1| > |2\eps_3\eps_5|$.

The matrix $U$ that diagonalizes $M$ via Eq.~(\ref{diag}) is given
approximately by
\beq
U \simeq \left( \begin{array}{cccc}
c_{01} & s_{01}e^{-i\delta_{01}} & 0 & 0
\\
-s_{01}e^{i\delta_{01}} & c_{01}
& {e^{i\phi_1}\over\sqrt2}(\eps_3e^{-i\phi_3}-\eps_5e^{i\delta_{23}})
& {e^{i\phi_1}\over\sqrt2}(\eps_3e^{-i(\phi_3+\delta_{23}}+\eps_5)
\\
\eps_3 s_{01} e^{i(\delta_{01}+\phi_3-\phi_1)}
&
-\eps_3 c_{01} e^{i(\phi_3-\phi_1)}
& {1\over\sqrt2} & {1\over\sqrt2} e^{-i\delta_{23}}
\\
\eps_5 s_{01} e^{i(\delta_{01}-\phi_1)}
&
-\eps_5 c_{01} e^{-i\phi_1}
& -{1\over\sqrt2}  e^{i\delta_{23}} & {1\over\sqrt2}
\\
\end{array} \right) \,,
\label{U}
\eeq
where
\bea
\tan \delta_{23} &=&
- {(\eps_4 - \eps_6)\sin\phi_1 + 2\eps_3\eps_5 s_3
\over (\eps_4 + \eps_6)\cos\phi_1 + 2\eps_3\eps_5 c_3 } \,,
\label{delta23app}
\\
\delta_{01} &=& \tan^{-1} \left({2\eps_3\eps_5 s_\alpha \over
\eps_1 - 2\eps_3\eps_5 c_\alpha} \right) - \phi_2 \,,
\label{delta01app}
\eea
with $c_\alpha \equiv \cos_\alpha$, $s_\alpha \equiv \sin_\alpha$,
$\alpha \equiv \phi_3 -\phi_1 - 2\phi_2$, and
\beq
\tan \theta_{01} = {\eps_2 \over \eps_1 + 2 \eps_3\eps_5}
\sqrt{1 + {4\eps_1\eps_3\eps_5(1-c_\alpha)
\over (\eps_1 - 2\eps_3\eps_5)^2} } \,.
\label{theta23app}
\eeq
We note that this $U$ has the form of of Eq.~(\ref{Ueps}). It also can
be seen to have the form of Eq.~(\ref{genU2}) if we set
\beq
\theta_{02} = \theta_{03} = 0 \,, \qquad \theta_{23} = {\pi\over4} \,,
\label{set}
\eeq
and make the identifications
\bea
s_{13} e^{-i\delta_{13}} &=&
{1\over\sqrt2} (\eps_3 e^{-i\phi_3} - \eps_5 e^{i\delta_{23}}) e^{i\phi_1} \,,
\label{ident1}
\\
s_{12} e^{-i\delta_{12}} &=&
{1\over\sqrt2} (\eps_3 e^{-i(\phi_3+\delta_{23})} + \eps_5) e^{i\phi_1} \,.
\label{ident2}
\eea

\begin{table}
\caption{\label{params} Parameters in the four-neutrino mixing matrix
and the primary observables used to determine them.}

\bigskip
\centering
\begin{tabular}{|c|c|}
& Primary \\
Parameter(s) & Observable(s) \\
\hline
$\theta_{01}$ & $A^{ee}_{sun}$ \\
\hline
$\theta_{23}$ & $A^{\mu y}_{atm}$ \\
\hline
$\theta_{12}$,$\theta_{13}$,$\phi_1\equiv\delta_{13}-\delta_{12}-\delta_{23}$
& $A^{e\mu}_{LSND}$,$A^{e\mu}_{atm}$,$B^{e\mu}_{atm}$ \\
\hline
$\theta_{02}$,$\theta_{03}$,$\phi_0\equiv\delta_{03}-\delta_{02}-\delta_{23}$
& $A^{\mu x}_{LSND}$,$A^{\mu x}_{atm}$,$B^{\mu x}_{atm}$
\end{tabular}
\end{table}

\begin{table}
\caption{\label{models} Summary of some particular four-neutrino models
for $m_0, m_1 \ll m_2 < m_3$ and $s_{02}, s_{03}, s_{12}, s_{13}
\sim \eps$. All models in the table have been constructed to have
short-baseline $\nu_\mu \rightarrow \nu_e$ ocsillations in agreement
with the LSND data and large-amplitude $\nu_\mu \rightarrow \nu_y$
oscillations in atmospheric and long-baseline experiments; they also
all have negligible $\nu_e \rightarrow \nu_x$ and $\nu_\mu \rightarrow
\nu_y$ oscillations in short-baseline experiments. The size of $\nu_\mu
\rightarrow \nu_x$ oscillations and $CP$ violation in long-baseline
$\nu_\mu \rightarrow \nu_y$ oscillations depend on other model
parameters. In all cases, one of $\nu_x$ and $\nu_y$ could be $\nu_\tau$
and the other sterile, or both could be sterile.}

\bigskip
\centering
\begin{tabular}{|ccc|ccc|c|}
&&& CP-conserving & CP-conserving & CP-violating & \\
&&& short-baseline & long-baseline & long-baseline & \\
Texture & $\delta_{23}$ & $\theta_{23}$ & $\nu_e \rightarrow \nu_y$
& $\nu_e \rightarrow \nu_\mu$ & $\nu_e \rightarrow \nu_\mu, \nu_y$
& Reference \\
\hline
$M_{e\mu}=0$ & ${\pi\over2}$ & ${\pi\over4}$ & No & Yes & No &
Ref.~\cite{bpww} \\
& ${\pi\over2}$ & $\ne {\pi\over4}$ & Yes & Yes & No &
Ref.~\cite{gmnr98} \\
& $\ne {\pi\over2}$ & any & Yes & Maybe & Yes &
Sec.~IV.A \\
& ${\pi\over4}$ & ${\pi\over4}$ & Yes & No & Maximal &
Eq.~(\ref{Mmaxmix}) \\
\hline
$M_{ey}=0$ & $0$ & ${\pi\over4}$ & Yes & No & No &
Ref.~\cite{mohap} \\
& $\ne 0,{\pi\over2},\pi$ & $\ne {\pi\over2}$ & Yes & Maybe & Yes &
Sec.~IV.B \\
& ${\pi\over4}$ & ${\pi\over4}$ & Yes & No & Maximal &
Eq.~(\ref{Mmaxmix}) \\
\hline
$M_{e\mu}, M_{ey} \ne 0$ & varies & ${\pi\over4}$ & Yes & Yes & Yes &
Eq.~(\ref{M}) \\
\hline
$M_{ee} = M_{e\mu} = M_{\mu\mu} = 0$ & ${\pi\over2}$ & ${\pi\over4}$ &
No & Yes & No & Sec.~IV.D
\end{tabular}
\end{table}

\end{document}